\documentclass[twocolumn]{aa}
\usepackage{graphicx}
\usepackage{natbib}

\newcommand{\COo}{\mbox{CO($J$=1--0)}}

\newcommand{\nine}{\times 10^9~{\rm M}_\odot}            %---"--- text

\begin{document}
   \title{Molecular gas in the galaxy M\,83
   \thanks{Based on observations collected with the Swedish-ESO Submillimetre 
Telescope at 
the European Southern Observatory, La Silla, Chile}}
   \subtitle{II. Kinematics of the molecular gas}

\author{A.~A.~Lundgren\inst{1,2} 
\and H.~Olofsson\inst{1} 
\and T.~Wiklind\inst{3,4}
\and G.~Rydbeck\inst{4}}

\institute{Stockholm Observatory, AlbaNova, SE-106 91 Stockholm, Sweden
\and European Southern Observatory, Casilla 19001, Santiago 19, Chile
\and ESA Space Telescpe Division, STScI, 3700 San Martin Drive Baltimore, MD 21218, USA
\and Onsala Space Observatory, SE-43992 Onsala, Sweden}

\offprints{Andreas Andersson Lundgren (andreas@astro.su.se)}

\date{\today}

\abstract{
We present the kinematics of the molecular gas in the barred spiral 
galaxy M\,83 (NGC5236). The study is based on $^{12}$CO(\mbox{$J$=1--0} and 
\mbox{2--1}) observations with the Swedish-ESO Submillimetre Telescope (SEST). 
Iso-velocity maps of the entire optical disk, 
10$\arcmin$\,$\times$\,10$\arcmin$ or 13\,$\times$\,13\,kpc, are 
produced. They show the pattern of an inclined, rotating disk, but also
the effects of streaming motions along the spiral arms. A dynamical 
mass of about 6\,$\times$\,10$^{10}$\,M$_\odot$ 
is estimated by fitting the rotation curve of an exponential disk 
model to these data. The 
gas constitutes about 13\% of the disk mass. 
The pattern speed is determined from the residual velocity pattern
The locations of various resonances are discussed. The molecular gas 
velocity dispersion is determined, and a trend of decreasing dispersion 
with increasing galactocentric radius is found. A total gas 
(H$_2$+\ion{H}{i}+He) mass surface density map is presented, and compared to 
the critical density for star formation of an isothermal gaseous disk. 
The critical density is exceeded in the spiral arms, but not in the 
interarm regions. The locations of Giant Molecular Associations 
(GMAs) and \ion{H}{ii} regions are consistent with this scenario of
dynamically induced star formation.
}

\maketitle

\section{Introduction} 
The kinematics of the gas in a galaxy provides fundamental clues to its
dynamics and evolution.
An excellent probe of the kinematics of disk galaxies is the molecular
gas (as traced by e.g.~CO radio line emission).
Since the molecular gas is collisional, the random motions are reduced
and the disk is dynamically cooled.
Because of this the molecular gas is able to respond, both quickly 
and strongly, to changes in the gravitational potential.
Atomic hydrogen is also a tracer which is often used, but since the 
scale height of \ion{H}{i} is relatively large compared to that of
the molecular gas it is less sensitive to the dynamics in the disk,
and more affected by forces not confined to 
the disk plane (e.g., shock waves from supernova explosions). 
Also, the centers of galaxies usually show very little \ion{H}{i} emission. 
Starforming regions give rise to a number of emission lines in
the optical regime that
can be used to trace the motion of the disk.
However, since the kinematics of these regions are affected by the 
starformation process it can be difficult to interpret such data.
Also, the high extinction in starforming regions further complicates matter.

The molecular gas kinematics is also an important probe of the
star formation process. A spiral density wave is an efficient mechanism 
for concentrating
atomic and molecular hydrogen in spiral features, but in order 
to initiate the star formation, the disk also needs to be
compressed in the vertical direction. In this smaller-scale process, 
dissipative processes like cloud-cloud collisions, interaction
with the galactic magnetic field, and the Jeans instability help enhancing 
the density.
Once a certain threshold density is reached, massive star formation
is initiated since the velocity dispersion no longer can neutralize
the self-gravity of the gaseous disk \citep{T64,K89}.

Two large surveys of the kinematics of spiral galaxies using CO radio line
emission have recently been completed: ``BIMA SONG'' done with the BIMA 
interferometer 
\citep{RTH01,HTR03} and ``The Virgo High-Resolution CO Survey''
done with the Nobeyama interferometer \citep{SKN03}. Both surveys
noted that some galaxies have a ``twin-peaks'' distribution of the
molecular gas in the center, and it is suggested that this is a result
of molecular gas agglomeration at the inner Lindblad resonance. More
results on the kinematics will come in subsequent papers from these
groups.
Many other galaxies have been observed in CO with interferometers and among
the most detailed studies of the global kinematics are studies done on
M\,100 with the IRAM interferometer \citep{GSC98} and one on M\,51 with
the OVRO interferometer \citep{AHS99}. 
In M\,100 the kinematics of the molecular gas
was modeled using a hydrodynamic code, and it was shown that the stellar bar
and the spiral pattern have a common pattern speed, while the nuclear
bar is rotating significantly faster. In M\,51 the authors found
streaming motions of very high velocities (60--150 km s$^{-1}$) and that two pattern
speeds coexist - one main mode and a second due to the
interaction with NGC\,5195.

Detailed studies done with single-dish telescopes of the dynamics of nearby 
galaxies with low inclinations are rare. The most well-studied galaxy is
M\,51, which has been observed in CO radio line emission with the Onsala 20m \citep{RHR85}, 
the IRAM 30m \citep{GCG93}, and the Nobeyama 45m telescope 
\citep{KNH95,KN97}. The last group showed that the gaseous disk is
gravitationally unstable in the arms, but not in the inter-arm regions,
and that the deviations from circular motion seen in the gas are in 
accordance with density-wave theory.

The kinematics of M\,83 has been observed using many tracers.
In particular, CO radio line data results have been presented by several 
authors; 
see \citet{LWO04} (Paper I) for a summary of CO observations of M\,83.
\citet{CEL78} produced the first CO rotation curve based on data in 7 positions.
Position-velocity diagrams of the nuclear region along the minor and major axes have
been published by \citet{HNS90}, \citet{IB01}, and \citet{DNT01}.
\citet{PW98} presented channel maps of the central 
$40\arcsec\times40\arcsec$, and discussed inner Lindblad resonances
and gas flow along the bar. \citet{LK91} used the
OVRO interferometer
to observe the eastern arm of M\,83 in the CO($J$=1--0) line. They presented channel 
maps and discussed the relation between CO, H$\alpha$, and \ion{H}{i}. The same
authors also observed the western bar end \citep{KL91}, and
they discussed, among other things, orbit crowding.
\citet{RLH99} revisited the eastern arm, extended the map, and
discussed streaming motions and disk stability.
The most extensive work on the molecular gas kinematics prior to this
paper was done by Crosthwaite et al. (2002), who mapped the disk in
CO(\mbox{$J$=1--0} and \mbox{$J$=2--1}) and presented a CO rotation curve, 
iso-velocity
patterns, and velocity dispersion patterns covering the optical disk.

The velocity pattern in \ion{H}{i}, both in the disk and in the \ion{H}{i} envelope,
and discussions of an \ion{H}{i} warp have been presented in papers
based on both interferometer data \citep{RLW74} and single-dish 
data \citep{HB81}.
\citet{TA93} compared velocity information obtained from \ion{H}{i} and H$\beta$ 
data, and discussed shocks, large scale H$_2$ dissociation, and the relation between
the dust and \ion{H}{i}.

\citet{C81} and \citet{DPD83} presented Fabry-Perot H$\alpha$ 
observations of M\,83. The latter group also discussed the pattern speed, 
the mass distribution, and the residual velocities.

In this paper, we present the kinematics of the molecular
gas, as seen in the CO(\mbox{$J$=1--0}) and  CO(\mbox{$J$=2--1}) 
lines, in the galaxy M\,83.
Two-dimensional velocity distribution and velocity 
dispersion maps are shown, and a rotation curve based on the assumption
of an exponential disk mass distribution is derived.
By subtracting the circular velocities
from the observed data we get the residual velocities.
Lastly, the location and formation of GMAs and the
stability of the gaseous disk are discussed.

\section{Observations} \label{observations}

Our CO($J$=1--0 and 2--1) observations were presented in Paper~I.
The data cover the entire optical disk, and is sampled at intervals
of 7$\arcsec$--11$\arcsec$. The spacing is small enough to facilitate
the use of a {\sc mem}-deconvolution technique. 
The data grid is oriented along the equatorial coordinate system and
centered on the coordinates $13^{\rm h}36^{\rm m}59\fs4$, 
$-29\degr52\arcmin05\arcsec$, which is the optical center given
by \cite{DDC76}. 
Since the grid spacing is small relative to the beam size, we can gain in S/N 
by convolving the raw data sets, without significantly affecting the angular 
resolution.
The results in this paper are therefore based on 
convolved CO(\mbox{$J$=1--0}) data (convolution kernel FWHM 20$\arcsec$, 
final angular resolution 49$\arcsec$),
convolved CO(\mbox{$J$=2--1}) data (kernel 15$\arcsec$, resolution 27$\arcsec$)
and two {\sc mem}-deconvolved CO(\mbox{$J$=1--0}) and 
CO(\mbox{$J$=2--1}) data sets
with angular resolutions of 22$\arcsec$ and 14$\arcsec$, respectively.
In order to compare data at the same angular resolution, data cubes 
with angular resolutions of 49$\arcsec$ 
(based on the raw data sets) and 27$\arcsec$
(based on the {\sc mem}-deconvolved data sets) were produced. 
The CO(\mbox{$J$=2--1}) data
have been corrected for emission in the error beam.
For a detailed description of the data, see Paper I.

\begin{table}
\caption{General Parameters of M\,83.}
\label{general}
\begin{tabular}{ll}
\hline
\hline
Morphological Type$^a$	&	SAB(s)c		\\
IR center:$^b$		&			\\
R.A.~(J2000)		&$13^{\rm h}37^{\rm m}00\fs8$		\\
Decl.~(J2000)		&$-29\degr51\arcmin56\arcsec$ 	\\
LSR systemic velocity (opt)$^c$&	506 km s$^{-1}$		\\
Distance$^d$		&	4.5 Mpc	\\
Position angle$^c$	&	45\degr\\
Inclination$^f$		&	24\degr		\\
Holmberg diameter (D$_0$)$^e$&	14\farcm6		\\
$M_{\rm \ion{H}{i}}$$^e$		&7.7$\nine$\\
$M_{{\rm H}_2}$$^g$		&3.9$\nine$\\
\hline
\noalign{\smallskip}
\noalign{$^a$ \citet{DDC76};
$^b$ \citet{SW94}; 
$^c$ \citet{C81}; 
$^d$ \citet{TTS03}; 
$^e$ \citet{HB81}; 
$^f$ \citet{TJD79};
$^g$ \citet{LWO04}}
\end{tabular}
\end{table}

\section{Kinematics}\label{kinematics}

\subsection{Channel maps}
\begin{figure*}
	\resizebox{\hsize}{!}{\rotatebox{-90}{\includegraphics{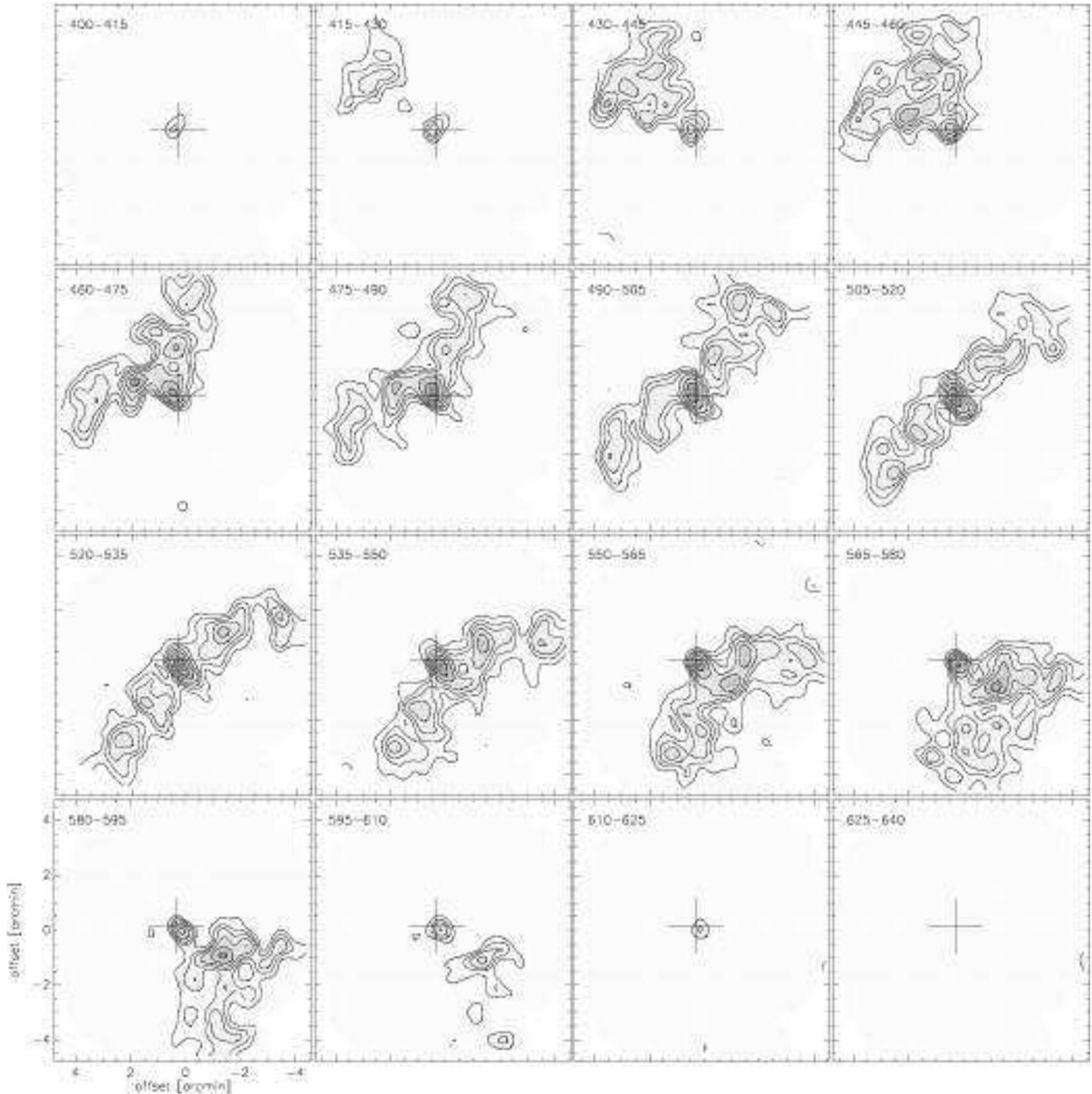}}}
	\caption{Channel maps of the {\sc mem}-deconvolved CO($J$=1--0) data set 
	ranging from 400 to 640 km s$^{-1}$ with a spacing
	of 15 km s$^{-1}$. The velocity interval is indicated 
	in the upper left corner of each panel.
	Contour levels are 0.5, 1.5, 3, 6, 12, and 24 K km s$^{-1}$. The
	cross marks the IR center.
	}
	\label{chanmaps}
\end{figure*}

In Fig.~\ref{chanmaps} channel maps ranging from 400 to 620 km s$^{-1}$ of 
the {\sc mem}-deconvolved CO(\mbox{$J$=1--0}) data cube, where each map covers a
velocity range of 15 km s$^{-1}$, are shown. They show the expected overall pattern
of an inclined, rotating disk. In addition, the arms are clearly resolved.
On the global scale the most notable deviation from the expected is
the outer arm on the NW side, which 
seems slightly offset from the expected velocity, an indication of streaming
motions or a warp. The prominent central region is visible
throughout the entire velocity range. The
emission peak moves from 
(+9\arcsec ,+2\arcsec ) to (-4\arcsec ,-8\arcsec )
relative to the IR center (See Table \ref{general}) -
a shift of 21\arcsec , which corresponds to 460 pc. These positions match
within a few arcseconds the two nuclear components found by 
\citet{PW98} in their CO(\mbox{$J$=3--2} and \mbox{4--3}) data.

\begin{figure*}
	\resizebox{\hsize}{!}{\rotatebox{-90}{\includegraphics{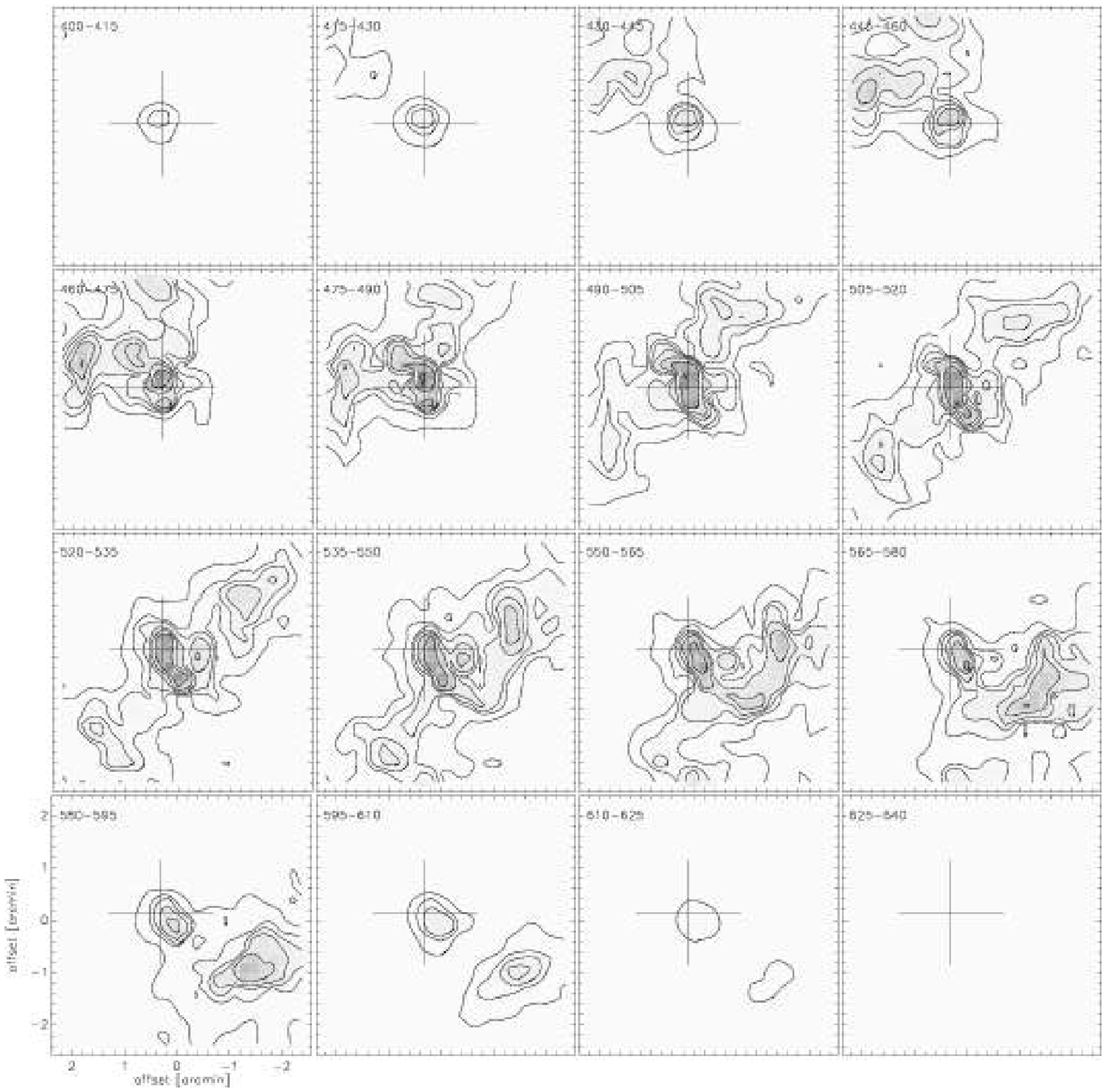}}}
	\caption{Channel maps of the central parts of the {\sc mem}-deconvolved 
	CO($J$=2--1) data set. The velocity interval and  
	contour levels are the same as in Fig.~\ref{chanmaps}.
	The cross marks the IR center.
	}
	\label{chanmaps21}
\end{figure*}

In Fig.~\ref{chanmaps21} the same channel maps in the 
{\sc mem}-deconvolved CO(\mbox{$J$=2--1}) data cube are shown. 
In order to highlight the
finer details provided by the better resolution in this data set, 
only the inner 140{\arcsec}$\times$140{\arcsec} are shown. Here, the
two central components are clearly resolved.

\subsection{Rotation curve and kinematic fit}
\label{secrotc}

There exist many procedures for
measuring the velocity of a spectral feature (see \citet{SR01} for
a recent review of methods). We chose the intensity-weighted-velocity
method to be able to directly
compare with results obtained using the same method applied to other velocity 
tracers, such as H$\alpha$ and \ion{H}{i}.
For a galaxy with a low inclination, like
M\,83, most methods give almost identical rotation curves. However,
in the nuclear region, where the velocity changes very fast with
respect to the beam size, it is likely that the intensity-weighted-velocity 
method underestimates the true rotational velocity. In Section \ref{pvma}
we also derive the rotation curve using the envelope-tracing
method \citep{STH97}.

In each spectrum the ``average'' velocity is estimated by calculating the 
intensity-weighted velocity. In order to minimize the effect of baseline 
variations, a sliding-window technique was used, where at each position 
integration was
performed inside a velocity window whose center and width depend on
the position within the galaxy. We checked carefully that the window
always covered all of the emission.

To obtain the best-fit circular velocities in the galaxy, the kinematic data 
have to be compensated for inclination, position angle, systemic velocity,
and kinematic-center offset. In order to estimate the values
of these parameters, a rotation curve is introduced
to which the velocity data sets can be compared.
Here we have chosen the theoretical rotation curve produced by an exponential
disk potential. The circular velocities in such a disk 
are given by:
\begin{equation}
v_{\rm c}^2(y)=2 G M_{\rm d} \frac{y^2}{R_{\rm d}}
	\left[ I_0(y)K_0(y)-I_1(y)K_1(y) \right],
\end{equation}
where $M_{\rm d}$ is the total mass of the exponential disk, 
$R_{\rm d}$ the disk scale length, and $y=R/2R_{\rm d}$ \citep{BT87}.
$I_{\rm n}$ and $K_{\rm n}$ are the modified Bessel functions of the
first and second kind, respectively.

We expect that the gas will not move in circular orbits in the bar region. 
Therefore, data obtained within 1\arcmin~of the nucleus were
excluded in the fit.
In order to minimize the influence of radial motions data 
from areas further away than $40\degr$ from the major axis were also excluded.
With this method it is not possible to determine the inclination angle, 
since the unique effects of variations of this parameter are seen as a 
changing opening angle 
of the iso-velocity contours along the minor axis\footnote{Along the 
major axis, an increase of inclination results in  a higher rotation 
curve, which can also be interpreted as a larger disk 
mass -- thus a degenerate problem.}. 
We adopted an inclination angle of 24$\degr$, which is in agreement 
with both the \ion{H}{i} iso-velocity contours and optical isophotes \citep{C81}.
All other parameters were fitted simultaneously to the kinematic data
sets obtained from the convolved CO(\mbox{$J$=1--0}) and 
CO(\mbox{$J$=2--1}) data cubes.
The fitting errors were minimized using a Levenberg-Marquardt method \citep{NUMREC}. 
We found that the kinematic center lies about $12\arcsec$ SE of the IR center 
(see Table \ref{general}), but within the errors of the fit the two
positions agree. 
The rotation curves derived for the south-western and north-eastern 
quadrants separately differed markedly in the inner 1\arcmin . 
This discrepancy disappeared when the kinematic offset 
parameters were fixed to match the location of the IR center.
We fitted the remaining parameters, and
the best-fit values (with errors) are given in Table \ref{parfit}. 
The same procedure was applied to the {\sc mem}-deconvolved data sets, 
and the same results were obtained (within the errors).
Our results agree closely with those obtained from the \ion{H}{i} data by 
\citet{TA93} (hereafter TA).

\begin{table*}
\caption{Best-fit parameters for our rotating-disk model. The kinematic
center was kept fixed to the IR center (coordinates given in Table 
\ref{general}). The data used are those of the convolved data sets.
}
\label{parfit}
\begin{tabular}{lllll}
\hline
\hline
				&\multicolumn{2}{c}{CO($J$=1--0)}				&\multicolumn{2}{c}{CO($J$=2--1)}			\\
				&Value				&Error (1$\sigma$)	&Value				&Error (1$\sigma$)\\
\hline
Kinematic center		&IR center	&fixed		&IR center	&fixed \\
Position angle			&$46\degr$	&$5\degr$	&$46\degr$	&$5\degr$\\
Inclination			&$24\degr$	&fixed		&$24\degr$	& fixed \\
Systemic vel. (LSR) [km s$^{-1}$] &$511.8$     	&0.6		&$511.5$	&0.7   \\
Disk mass  [$10^10$ M$_{\odot}$]&5.9		&0.7		&6.2		&0.8\\
Disk scale length [kpc]    	&2.7  		&0.2    	&2.9		&0.2\\
\hline
\end{tabular}
\end{table*}

The rotation curve fit results in a dynamical mass of 6\,$\times$\,10$^{10}$ 
${\rm M}_\odot$. 
(\citet{CTB02} derived a dynamical mass of 8\,$\times$\,10$^{10}$ 
${\rm M}_\odot$ (for a distance of 4.5 Mpc)
using a Brandt model rotation curve.)
The total mass of molecular and atomic hydrogen gas within the Holmberg
limit ($5.5\nine$, see Paper~I) constitutes 
about 9\%$\pm$2\%  of the estimated disk mass. 
Including helium the total gas mass is about 13\% of the disk mass.
This fraction depends somewhat on the choice of the inclination angle
e.g., changing the inclination from
24 to 26 degrees increases the hydrogen gas fraction from 9\% to 10\%.

\begin{figure*}
	\centerline{\resizebox{0.8\hsize}{!}{\rotatebox{-90}{\includegraphics{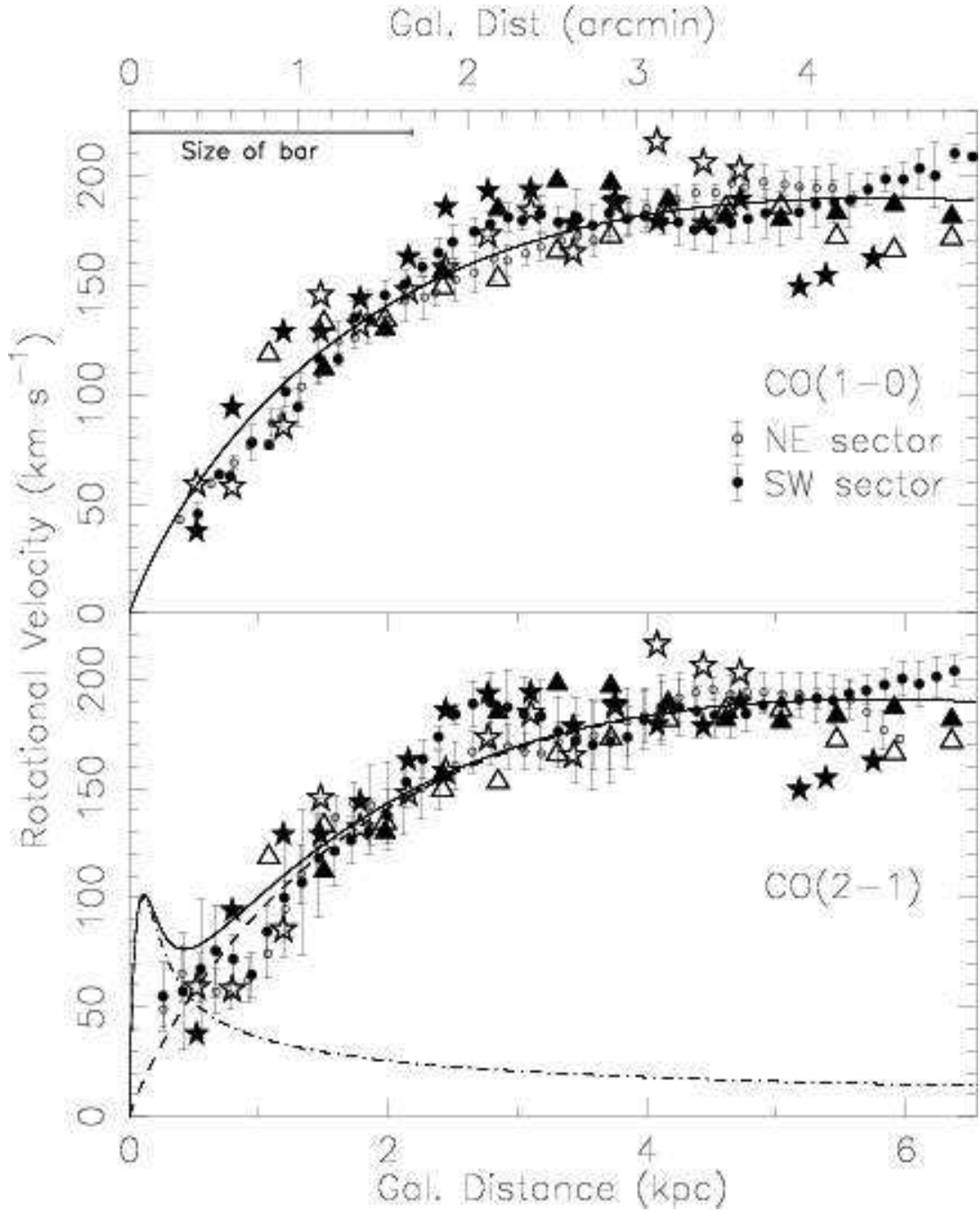}}}}
	\caption{Circular velocities for the convolved CO($J$=1--0) (upper panel)
	and {\sc mem}-deconvolved CO($J$=2--1) (lower panel) data sets (within
	$40\degr$ of the major axis, binned into 6\arcsec~(130 pc) 
	wide annuli) shown as circles. 
	Filled and unfilled circles represent data from the north-east 
	and south-west sector, respectively, and the error bars are 1$\sigma$.
	Circular velocities from H$\alpha$ [stars] (de Vaucouleurs et al.~1983) 
	and \ion{H}{i} [triangles] (TA) are also shown. 
	The solid lines are fits of a rotation curve generated by an 
	exponential disk (upper panel) and a disk + central component
	 (lower panel) model.}
	\label{rotc}
\end{figure*}

The circular velocities rise linearly (and slightly slower
than the model rotation curve) in the central regions
and level off at larger radii, upper panel in Fig.~\ref{rotc}.
In this figure we have for comparison also plotted circular velocities
from H$\alpha$ \citep{DPD83} and \ion{H}{i} (TA). 
Inspection of position-velocity diagrams along the major
axis (Fig.~\ref{posvel}) indicates that there are substantial
deviations from the rotation curve due to e.g.~streaming motions (see
Sect.~\ref{secvelres}).

Although not readily visible from the azimuthally-averaged rotation
curve, a central component of relatively low mass may exist.
Indeed, the position-velocity diagrams along the major 
axis, shown in Fig.~\ref{posvel}, suggest a more rapidly rising rotation
curve with a first maximum at $\approx$\,15$\arcsec$.
Adopting, as an example (and consistent with our estimate
from the position-velocity diagrams  in Sect.~\ref{pvma}), 
an inner (nuclear) exponential 
disk\footnote{The type of mass distribution
has relatively little effect on the behavior of the rotation curve
as long as the scale length of the mass distribution is sufficiently
small.} with a mass of $3\times10^8 \ {\mathrm M_\odot}$,  
and a disk scale 
length of 50 pc, and plotting the new rotation curve together with 
the circular velocities
obtained from the deconvolved CO(\mbox{$J$=2--1}) data set (lower panel in 
Fig.~\ref{rotc}), it is clear that the effects of such a mass distribution 
can very well be hidden in our data.
(The mass and the scale length of the outer disk
remain the same.)

\subsection{Resonances}
\label{secreso}
Objects orbiting the center of a galaxy oscillates radially
in addition to their motion in circular or nearly-circular 
orbits. 
If the frequency of these radial oscillations, called the epicycle 
frequency ($\kappa$), is commensurate with the angular velocity
of the bar and spiral arms, resonances occur.
Depending on the conditions, the resonance can trap objects in a 
density wave or deplete a region. 
These resonances occur when $\Omega=\Omega_p$ (corotation, CR) or when
$nm(\Omega-\Omega_p)=\pm\kappa$, 
where $m$ is the number of arms (in this case 2), $n$ an integer number $\ge$\,1, $\Omega$ the 
angular velocity of the gas and stars, and $\Omega_{p}$ the pattern 
speed. The case $n$\,=\,1 results in the Lindblad resonances, while $n$\,=\,2 corresponds
to the lowest order of the ultraharmonic resonances.
In barred galaxies it is generally assumed that the bar and the
spiral pattern rotate with a common pattern speed.
In the following discussion, we will assume that the bar and the 
spiral arms have a common pattern speed of 47 km s$^{-1}$ kpc$^{-1}$
(see Sect.~\ref{secvelres}).

\begin{figure*}
	\centerline{\resizebox{0.8\hsize}{!}{\rotatebox{-90}{\includegraphics{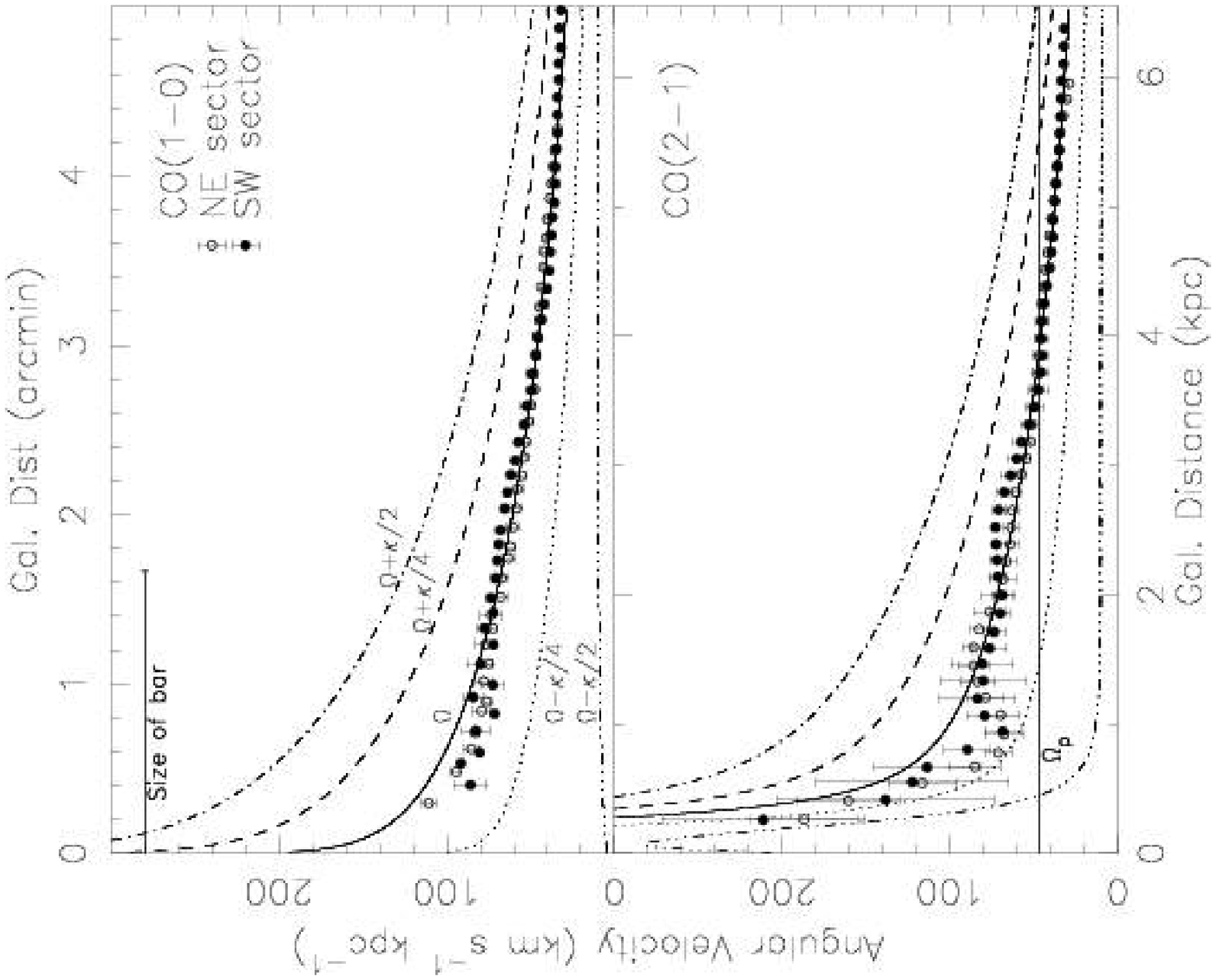}}}}
	\caption{The angular velocity as a function of radius. The 
	velocity data in the upper and lower panels are taken from the 
	CO($J$=1--0) convolved and the {\sc mem}-deconvolved CO($J$=2--1) data 
	sets, respectively.
	The solid line in the upper panel is the the best fit of the 
	adopted rotation curve to the data (see Table \ref{parfit}). 
	In the case of the CO($J$=2--1) data the rotation curve
	is composed of a disk and a central component (see Fig.~\ref{rotc}).
	Dashed and
	dotted lines represent: $\Omega-\kappa$/2, 
	$\Omega-\kappa$/4, $\Omega+\kappa$/4, and $\Omega+\kappa$/2, 
	where $\Omega$ is the angular velocity
	and $\kappa$ the epicyclic frequency. The thick solid line in the bottom
	panel shows the adopted pattern speed (47 km s$^{-1}$ kpc$^{-1}$).}
	\label{omega}
\end{figure*}

\begin{figure*}
	\resizebox{\hsize}{!}{\rotatebox{-90}{\includegraphics{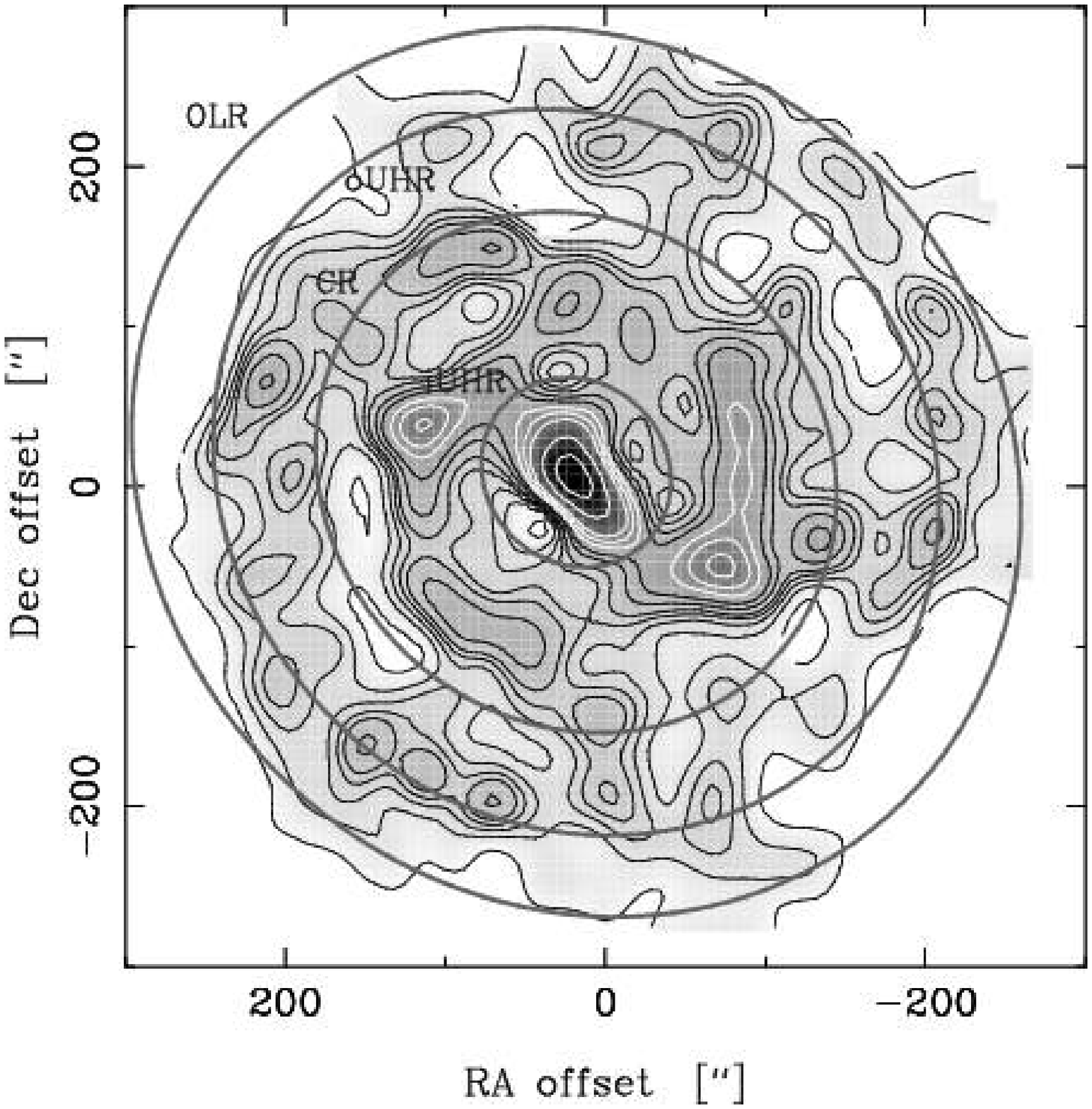}}}
	\caption{The radii where our model predicts the locations of the
	inner ultra harmonic
	resonance (iUHR), corotation (CR), outer ultra harmonic resonance
	(oUHR), and outer Lindblad resonance (OLR) in M\,83, on the 
	velocity-integrated CO(\mbox{$J$=1--0}) intensity in the {\sc mem}-deconvolved
	data set.}
	\label{crlin}
\end{figure*}

This results in a corotation radius (CR) of $\approx$170$\arcsec$ (3.7 kpc),
an outer UHR (oUHR) at $\approx$240$\arcsec$ (5.2 kpc),
and an outer Lindblad resonance (OLR) at $\approx$290$\arcsec$ (6.3 kpc),
Figs~\ref{omega} and \ref{crlin}.
The OLR corresponds to the scale of the optical and molecular gas disk. 

While introducing a low-mass nuclear component has a negligible effect
on the rotation curve, it has a noticeable effect on the
angular-velocity diagram. 
In the case with the single-disk mass distribution (solid line in the upper
panel in Fig.~\ref{omega}) 
there is no indication of an inner Lindblad resonance (ILR), which arises
where $\Omega-\kappa/2=\Omega_p$.
However, the CO emission at the nucleus of M\,83 is split into two distinct 
components (\cite{PW98}; Paper~I), and it has been shown that the location 
(and existence) of a double nucleus of molecular gas in barred galaxies, 
is coincident with the location of an ILR \citep{KWS92}.
In addition, \citet{ECW98} observed a nuclear ring in a $(J-K)$ 
color image of M\,83. The radius of this ring is 9\arcsec~and the authors
proposed that it coincides with the outer ILR (oILR).
The disk plus central component model does allow for the existence of an 
ILR (lower panel in Fig.~\ref{omega}). 
In the example above with an inner disk mass of 
$3\times10^8 \ {\mathrm M_\odot}$, the oILR would be located at 
a radius of $\approx 20\arcsec$. The possibility of an ILR is therefore
not inconsistent with our data, but its exact location cannot be determined
(nor can its existence with any certainty) due to the limited spatial resolution
and grid spacing of our data.

Figure~\ref{omega} indicates the presence of an inner Ultraharmonic Resonance 
(iUHR), where $\Omega-\kappa/4=\Omega_p$, at a radius of 
60$\arcsec \pm$20$\arcsec$ (1.3$\pm$0.4 kpc) (see also Fig.~\ref{crlin}).
(The distance from the nucleus to the iUHR is only about 5$\arcsec$ smaller if 
the low-mass nuclear component is not present.)
In this radial range in the SW part of the bar a very conspicuous
horn-shaped feature can be seen in both our CO($J$=2--1) map (Fig.~\ref{isovel21}) 
and in H$\alpha$ images. In the NE section of the bar there is an abrupt
decrease of the H$\alpha$ emission along the leading edge of the bar between the 
radii 45--55$\arcsec$, and in the CO($J$=2--1) map the emission shows a sharp bend 
at this location.
Furthermore, between radii 40$\arcsec$ and 60$\arcsec$, the CO($J$=2--1) intensity
drops by a factor of about 4, and the corresponding drop in the CO($J$=1--0)
intensity is about 2--3. 
It is possible that these features are the results of the
interaction between gas and stars on square-shaped 4:1 
orbits and material flowing along the leading edge of the bar towards
the nucleus. 
\citet{RSV99} offer another explanation: their SPH-modeling suggests that
some gas that initially follow the dust lane on the leading edge of the 
bar will over-shoot the nuclear region and end up in the ``spray region'' 
on the opposite side of the bar in a region behind the dust lane.
When this gas encounters gas flowing along the leading edge of the 
bar their trajectories will be more or less orthogonal to each other,
and the interaction could have a strong impact on the gas flow in the
dust lane, particularly if the gas feeding process is intermittent.

We note that the spiral pattern of M\,83 appears to avoid the
CR region. The spiral arms cross the CR region at (100$\arcsec$,150$\arcsec$) 
and (-80\arcsec$,-130\arcsec$) 
in the coordinate system of Fig.~\ref{crlin}. The pitch angle
increases abruptly when the arms approach the CR and decreases
again on the other side of the CR region.
Upon encountering the OLR, 
the pitch angle decreases and becomes negative. 
The spiral pattern shows here the
characteristic dimpled shape of the 2:1 orbits that run in the retrograde
sense in the rotating frame inside the OLR, e.g.,~\citet{K91}.

The decrease of the angular velocity 
around 40$\arcsec$ ($\approx$1 kpc) could be due to the effect of streaming motions
in the bar. Gas on $x_{1}$ orbits, i.e.~elongated orbits along the bar, 
will have 
a rotational velocity at the apex of the ellipse which is lower than
that of gas on a circular orbit at the same
radius. Hence, the $\Omega$ will be lower. 
In the SW data,
at galactocentric distances between 2 and 3 kpc, all measurements indicate 
deviations from circular motion. This can also be seen in the angular-velocity 
diagram: the NE sector shows a steady
decrease of the angular velocity as a function of radius, while the 
SW sector has a region where the curve is more or
less flat. This is the location of the western bar end, and we will
discuss a possible explanation for this behavior in the next section.

\subsection{Isovelocity maps and residual velocities}
\label{secvelres}
\begin{figure*}
	\resizebox{\hsize}{!}{\rotatebox{-90}{\includegraphics
	{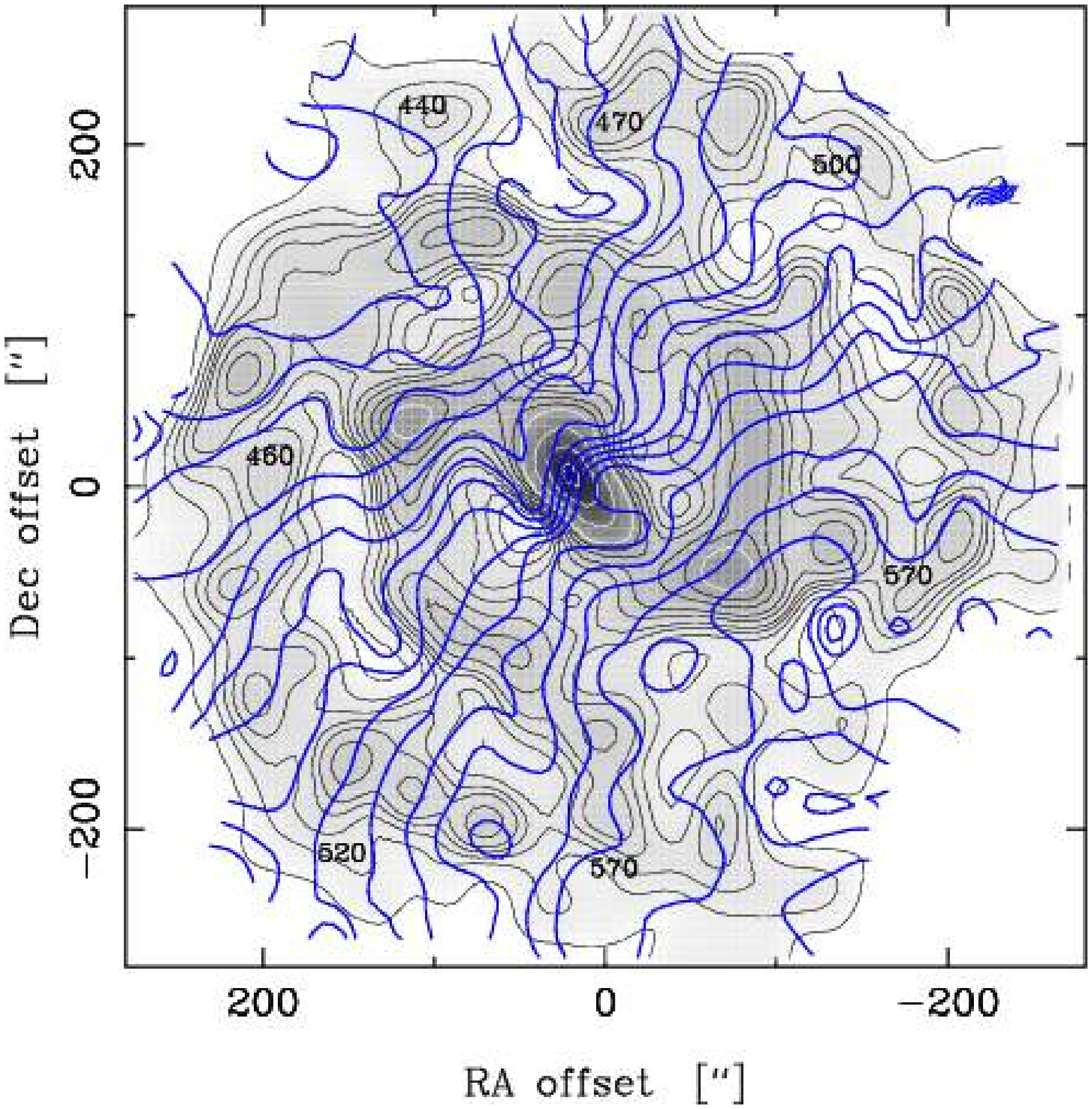}}}
	\caption{Isovelocity curves (black lines) superposed on the 
	velocity-integrated CO(\mbox{$J$=1--0}) intensity in the {\sc mem}-deconvolved
	data set (contour increment 10 km\,s$^{-1}$).}
	\label{isovel10}
\end{figure*}

\begin{figure*}
	\resizebox{\hsize}{!}{\rotatebox{-90}{\includegraphics
	{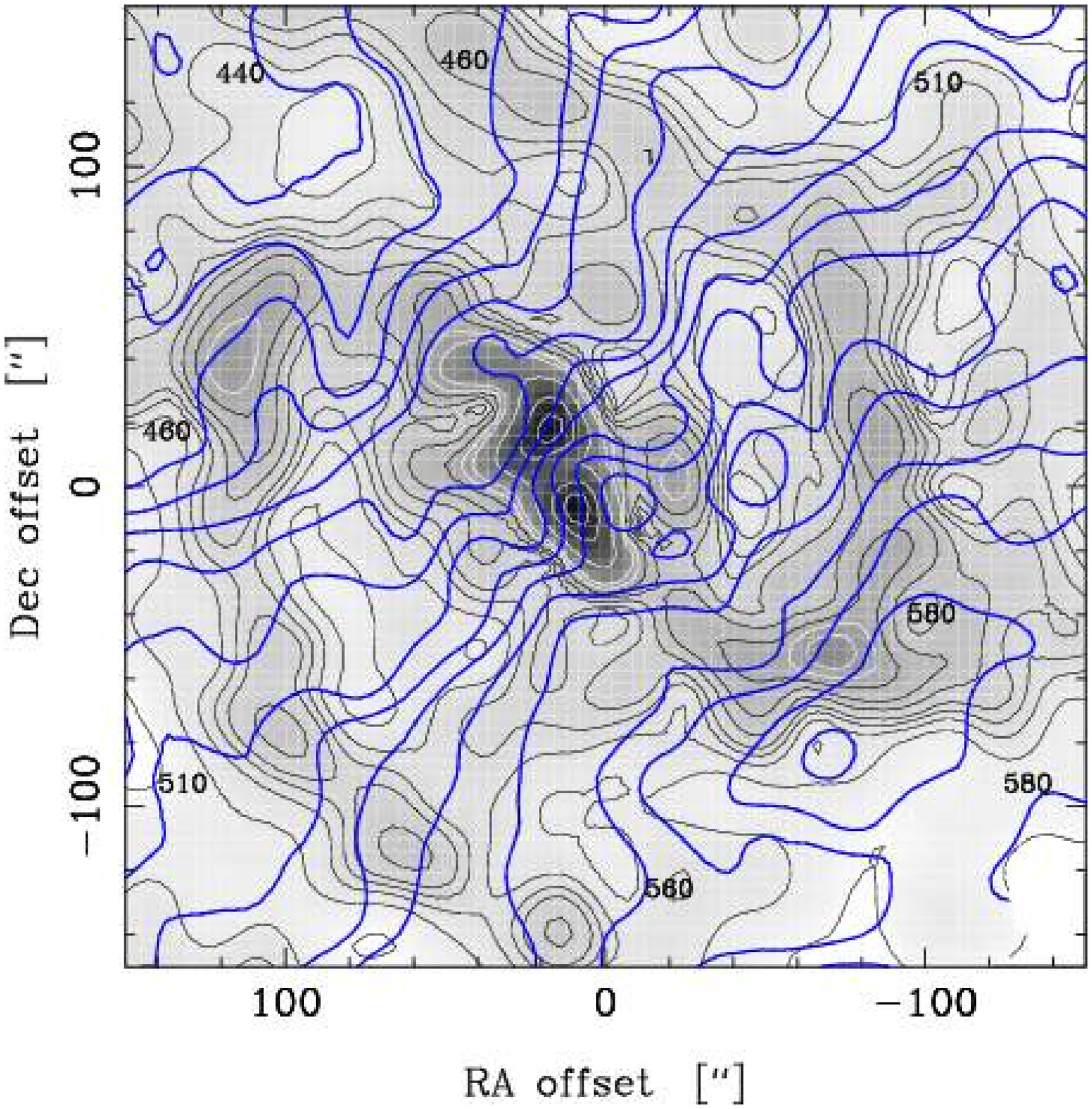}}}
	\caption{Isovelocity curves in the inner part of M\,83 superposed 
	on the velocity-integrated CO(\mbox{$J$=2--1}) intensity in the 
	{\sc mem}-deconvolved data set (contour increment 10 km\,s$^{-1}$).}
	\label{isovel21}
\end{figure*}

In Figs~\ref{isovel10} and \ref{isovel21} the isovelocity 
contours obtained from the deconvolved CO(\mbox{$J$=1--0}) and 
CO(\mbox{$J$=2--1}) 
data sets, respectively, are plotted on 
grey-scale maps of the CO line emission in the respective transition. 
Both maps show the characteristic pattern of an inclined, rotating disk.
Wiggles, indicating deviations from 
local circular motion, are present in the entire disk.
Along the leading edge of the bar (see Fig.~\ref{isovel21}), the shape
of the iso-velocity contours points towards lower circular velocities
than expected. This is caused by non-circular motions due
to the oval gravitational potential in the bar and/or a shock associated
with the leading edge of the bar.  
At the center of the galaxy the iso-velocity contours
are tilted by $45\degr$ with respect to the minor axis, suggesting strong
non-circular or out-of-plane motions (the former is expected in the presence
of a strong bar).
In CO we do not detect any strong evidence of the warp seen in \ion{H}{i} 
\citep{RLW74,HB81,TA93,CTB02}. This is not an unexpected result since a
warp  mainly affects regions outside the optical disk.

\begin{figure*}
	\resizebox{0.95\hsize}{!}{\rotatebox{0}
		{\includegraphics{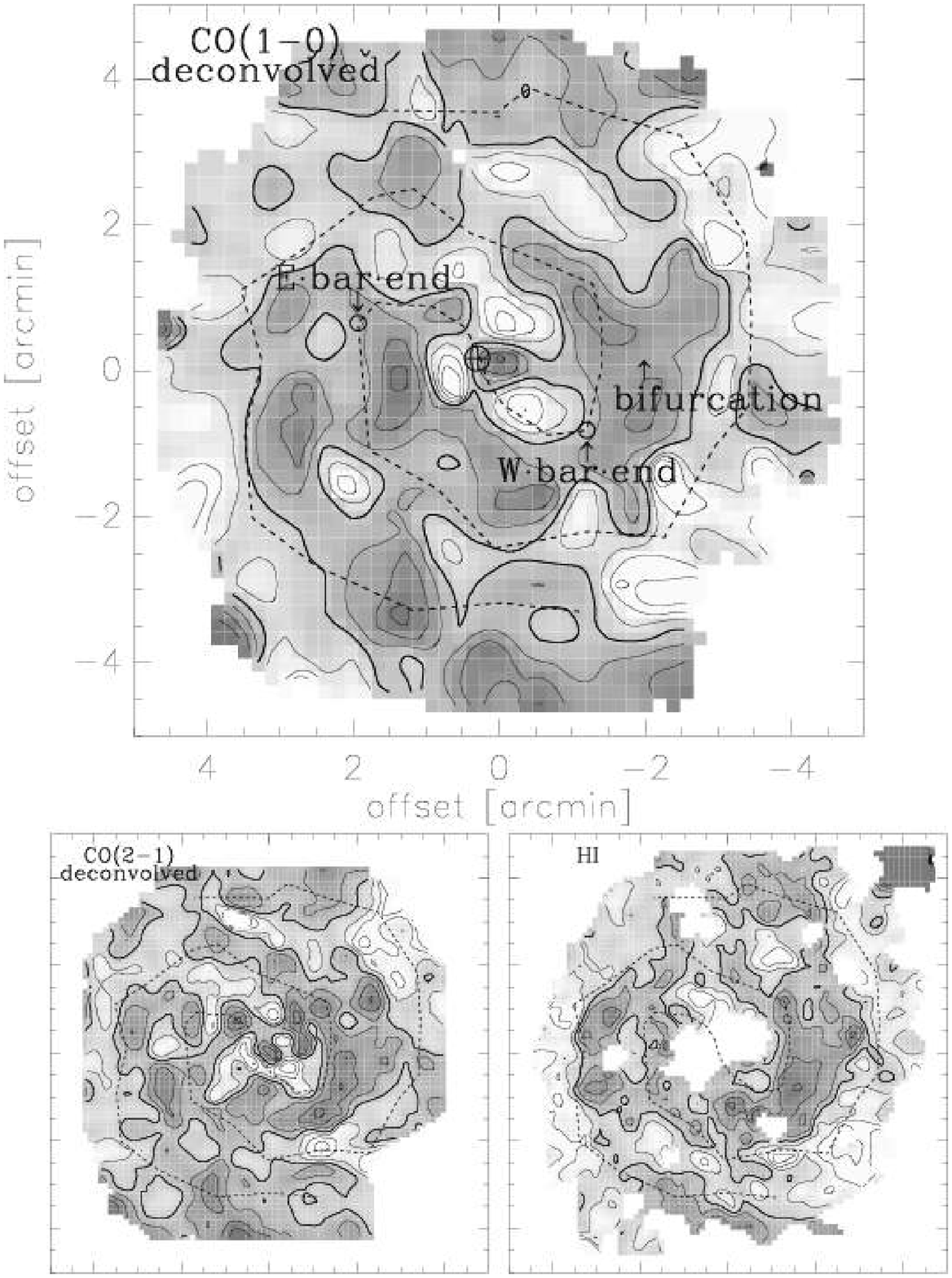}}}
	\caption{Residual-velocity maps after subtraction
	of the axi-symmetric component (described in the text)
	from the observed velocity fields in the deconvolved CO(\mbox{$J$=1--0}),
	deconvolved CO(\mbox{$J$=2--1}), and \ion{H}{i} data sets. The resolutions
	are 23\arcsec , 13\arcsec , and 12\arcsec , respectively.
 	The grey-scale and contour interval is $-15$ km\,s$^{-1}$ (white) 
	to +15 km\,s$^{-1}$ (dark).
	The contour increment is 5 km\,s$^{-1}$ and the thick
	contour represents 0 km\,s$^{-1}$.
	In the top image the position of the nucleus is shown as 
	an encircled cross, and the positions of the bar ends are marked
	with circles.
	In all images the spiral arms, as seen in CO radio line data, are 
	drawn with a dashed line.
	}
	\label{residuals}
\end{figure*}

Displayed in Fig.~\ref{residuals} are the residual velocities,
remaining after subtraction of the axi-symmetric-model velocity field,
in the {\sc mem}-deconvolved CO(\mbox{$J$=1--0}) and
CO(\mbox{$J$=2--1}) data sets, as well as the \ion{H}{i} data set from TA.
In order to have comparable residual maps we subtracted the same
axi-symmetric velocity field (the one obtained when fitting to the 
convolved CO(\mbox{$J$=1--0}) data set). Also shown in these figures is a line marking 
the crest of the gas mass surface density along the spiral pattern, 
as obtained from the CO(\mbox{$J$=1--0}) and CO(\mbox{$J$=2--1}) 
images. The residuals seem to follow a 
spiral-shaped pattern in all maps.
In general, the residual velocities in
the different maps show a good resemblance: both in the 
total range (from $-20$ to +20 km\,s$^{-1}$) and in 
specific details, e.g., the large region with positive residuals
3$\farcm$5~S of the center, and the bifurcation 2$\farcm$5~W of the center.
The \ion{H}{i} map has a slightly better resolution (12\farcs 4), 
but our maps show in addition the residuals in the central region
where the \ion{H}{i} emission is absent. 
Some regions in the residual velocity maps show very sharp gradients, 
such as the end of the arm bifurcation to the west. These areas can be 
identified in the \ion{H}{i} 
residual-velocity map as well. All of these regions coincide 
with areas in the disk with low mass surface density in both molecular 
and atomic hydrogen. Our interpretation is simply that the
relatively low gas mass surface density in these regions is due to
some mechanism, such as e.g.~``shock focusing'' \citep{LK91}, 
that regulates how much gas can enter the region, and in the
process skews the velocity distribution of the gas.

An interesting aspect of the residual-velocity pattern is that it may
be used to find the pattern speed. The technique is called
``geometric-phase technique'', and it is described in \citet{C93}. 
According to this method, the residual-velocity pattern  goes from being
a one-armed pattern to a three-armed pattern at the corotation radius 
in a two-armed
spiral galaxy, where the pattern is driven by a spiral density wave
(i.e., the pattern bifurcates at CR). 
On the western side there is such a
bifurcation, but on the eastern side the situation is less clear. 
The galactocentric distance to the bifurcation on the western side
is 160--180\arcsec~(3.5--3.9 kpc), i.e., at 1.6--1.8 $R_{\rm bar}$
($R_{\rm bar}$ is estimated as the radius where the isophotes in a K-band image
become significantly influenced by the spiral arms).
This is inconsistent with the results from early 
analytical calculations \citep{CG89}, which show that
corotation occurs just outside the end of the bar, and also larger than
results from numerical SPH simulations \citep{A92}, which suggest a distance to the CR of 1.2$\pm$0.2 $R_{\rm bar}$.
Using Fig.~\ref{omega} we deduce that the angular pattern speed 
of M\,83 is about 40--55 km\,s$^{-1}$\,kpc$^{-1}$. This is in excellent
agreement with the value 50$\pm$9 km\,s$^{-1}$\,kpc$^{-1}$ derived by
\citet{ZR03} by applying the Tremaine-Weinberg method
\citep{TW84} to our \COo\ data set. 

\subsection{Position-velocity diagrams}
\label{pvmaps}
The position-velocity (PV) diagrams are obtained along the position angles 
45\degr~(major axis, left row in Fig.~\ref{posvel}), and 
--45\degr~(minor axis, right row in Fig.~\ref{posvel})
across the IR center of the galaxy. 
We used the deconvolved CO(\mbox{$J$=1--0}) and CO(\mbox{$J$=2--1}) data sets,
but the behavior is identical in the convolved data sets, however, with
a lower angular resolution. We further assume in our interpretation
that all motions take 
place in the plane of the galaxy (inclination 24\degr ), and that the
NW side is the near side.

\subsubsection{Major axis} 
\label{pvma}

Along the major axis of the galaxy the CO(\mbox{$J$=1--0}) PV-diagram crosses
spiral arms twice, and there are velocity gradients over both arms. 
When compared to the expectations from density-wave-driven streaming motions 
in a disk with trailing arms, e.g., \citep{V80b,C93}, the effect is 
most pronounced in the SW arm.
Outside CR and outside the arm, the gas is slowed down in the azimuthal 
direction by the more rapidly moving spiral potential, while on the inside 
it is accelerated. 
In the PV-diagram this appears as 
an increase in velocity, above that of the rotation curve, before the arm and 
a decrease in velocity, below that of the rotation curve, beyond the arm 
(in the direction of increasing galactocentric radius; 
note that the exact location of the spiral potential minimum is somewhat uncertain
and not necessarily coincident with the CO arms which are indicated in the figure). 
The total azimuthal velocity
shift is about 15 km s$^{-1}$, or about 35 km s$^{-1}$ in the plane of the
galaxy.  This is comparable to the streaming motions observed in M\,81 
\citep{V80b,V80,AW96}, but smaller than the ones observed in M\,51 
\citep{RHR85,GGC93,KN97,AHS99}.
The situation is less clear-cut in the NE arm, but the behavior is not
inconsistent with that of density-wave-driven streaming motions. 
One explaination for the lack of expected streaming signatures at the NE crossing might be due 
to its proximity to the CR radius. Here the gas is comoving with the gravitational
potential and thus subject to strongly non-linear effects.

\begin{figure*}
	\resizebox{\hsize}{!}{\rotatebox{-90}{\includegraphics{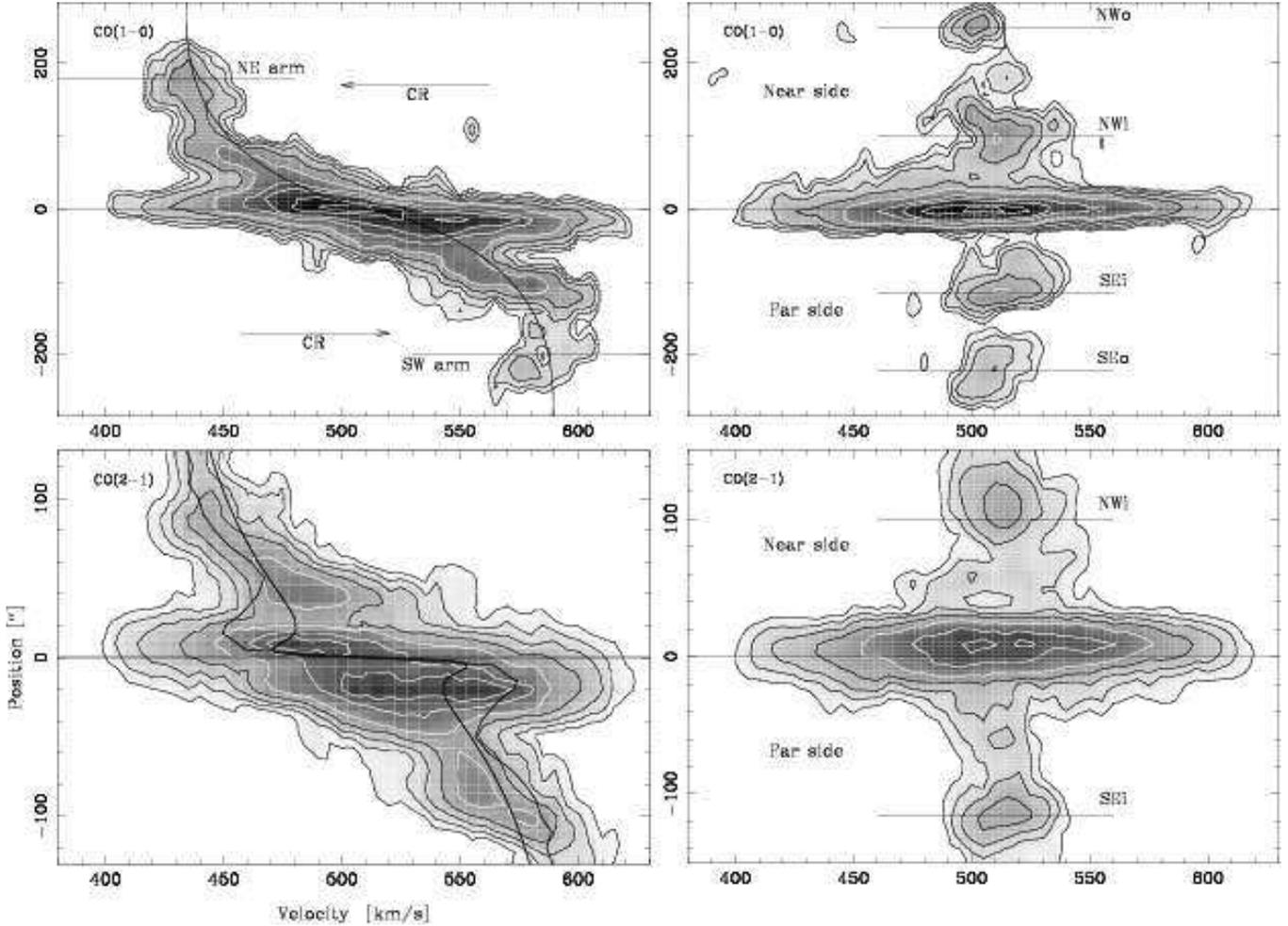}}}
	\caption{Position-velocity diagrams along the major (left column) and
	minor (right column) axes, across the IR center, in
	CO(\mbox{$J$=1--0}) (upper panels) and 
	CO(\mbox{$J$=2--1}) (lower panels) emission.
	The contours are 0.035, 0.07, 0.14, 0.3, 0.6, 1.2, 
	1.8, 2.4 and 2.8\,K (first white contour is 0.6\,K).
	In order to show the details at the center, the
	y-scale is expanded in the CO(\mbox{$J$=2--1}) panels.
	The rotation curves obtained in Sect.~\ref{secrotc} are drawn as solid
	lines in the major axis diagrams.
	Additionally, in the lower left panel, the rotation curve obtained
	with the envelope-tracing method is also drawn. The latter rotation curve
	can be identified by its systematically higher higher rotational
	velocities.
	}
	\label{posvel}
\end{figure*}

In the central region the velocity changes 
by about 50--75 km\,s$^{-1}$ over $\approx$30$\arcsec$ when crossing from 
the SW to the NE side of the nucleus
(the effect is most pronounced in the CO(\mbox{$J$=2--1}) map).
If this motion lies in the plane of the galaxy, the rotation speed 
is 60--90  km\,s$^{-1}$ at a distance of 15$\arcsec$ (330 pc) from the nucleus.
This suggests a much more rapidly rising rotation curve in the central region 
than the one obtained from the azimuthally-averaged data in 
Sect.~\ref{secrotc}, and the presence
of a central mass component (as discussed in Sects~\ref{secrotc} and 
\ref{secreso}).
Assuming a rotational speed of 60 km\,s$^{-1}$, the enclosed
mass within 330 pc is about $3\times10^8 \ {\mathrm M_\odot}$.
The emission in this region is dominated by the two central components 
observed in e.g.~CO(\mbox{$J$=3--2} and \mbox{$J$=4--3}) by \citet{PW98}, 
and in CO(\mbox{$J$=2--1}) by us (Paper~I). Note that the velocities of these 
concentrations, about 480 and 555 km\,s$^{-1}$ in the CO(\mbox{$J$=2--1}) 
map, are not symmetrical with respect to the systemic velocity,
$V_{\rm sys}$=510 km\,s$^{-1}$). 
This can  be explained by the fact that the two central components 
lie at different radial distances from the kinematical center.
The result is also consistent with the CO(\mbox{$J$=1--0}) PV-diagram 
of \citet{HNS90}, and
the higher-$J$ CO line PV-diagrams of \citet{IB01} and \citet{DNT01}. 

In the lower left panel of Fig.~\ref{posvel} we also show the 
rotation curve obtained using the envelope-tracing method to estimate
velocities in a major-axis PV diagram \citep{STH97}. The first step is to
locate the terminal velocity ($v_{\mathrm t}$), i.e., the velocity at which the intensity 
is 20\% of the peak intensity, for each radius (see e.g.~\citet{SR01} 
for details of the method). The circular velocity ($v_{\mathrm c}$) is then 
calculated using
\begin{equation}
	v_{\mathrm c}=(v_{\mathrm t}-v_{\mathrm {sys}})/\sin i - 
	(\sigma_{\mathrm {obs}}^2 + \sigma_{\mathrm {ISM}}^2)^{1/2}
\end{equation}
where $\sigma_{\mathrm {ISM}}$ is the
velocity dispersion of the ISM, and $\sigma_{\mathrm {obs}}$ is the
velocity resolution of the observations.
We used the radial distribution of the total velocity dispersion shown
in Fig.~\ref{specwidth} as an estimate of $(\sigma_{\mathrm {obs}}^2 + \sigma_{\mathrm {ISM}}^2)^{1/2}$.
The circular velocities obtained in this way are systematically higher
than the ones obtained with the intensity-weighted-velocity 
method. The shape of the rotation curve is consistent with a massive central
component. 
An even steeper rotation curve has been derived by \citet{STH99} using this 
method on \ion{H}{i} and CO data.
The main reason for the difference lies in the higher spatial resolution
of the latter observations. Finally, it should be noted that this rotation
curve is obtained along the bar, which happens to lie along the major axis.
In this region there may also be non-negligable motions perpendicular to
the plane.

\subsubsection{Minor axis}
Along the minor axis the  CO(\mbox{$J$=1--0}) PV-diagram crosses 
the arms four times; denoted, starting from lower left, SEo (South-East outer), 
SEi (South-East inner), NWi, and  NWo.

In the SEi and SEo crossings the behavior is once again as expected 
from density-wave-driven streaming motions. The gas is accelerated 
radially outwards inside the arms, and radially  inwards 
outside the arms. The velocity shifts are about 15 km s$^{-1}$, corresponding to
about 35 km s$^{-1}$ in the plane of the galaxy, i.e., consistent with
the velocity shift estimated at the major axis NE crossing. In the NW part
the situation is much more complicated. The NWo crossing may be affected by a
warp of the plane as suggested by \ion{H}{i} data \citep{TA93}. Between the NWo 
and NWi crossings there
is gas with radial velocities approaching 75 km s$^{-1}$ in the plane of the galaxy,
both inwards
and outwards, and this makes the interpretation of the NWi crossing more difficult.
The gas moving inwards comes from an area slightly east of the minor axis, 
in a region where the arm bends sharply. Presumably this has an effect on
the kinematics. In general, the gas is much more perturbed in the NW part. 

\subsection{Velocity dispersion}
\label{secveldisp}

We estimate the velocity dispersion in the spectra by measuring
their equivalent width, i.e., the velocity-integrated intensity 
divided by the peak intensity. This is converted to 
the standard, intensity-weighted, velocity dispersion, $\sigma_{\rm CO}$,
by dividing the result with $\sqrt{2\pi}$, i.e., the conversion constant
for a Gaussian profile.
Prior to the calculation of the velocity dispersions 
the spectra were smoothed to 5 km\,s$^{-1}$ in order to minimize 
the impact of the noise. This also facilitates the comparison of data
with different velocity resolutions.

\begin{figure*}
	\resizebox{0.48\hsize}{!}{\rotatebox{-90}
		{\includegraphics{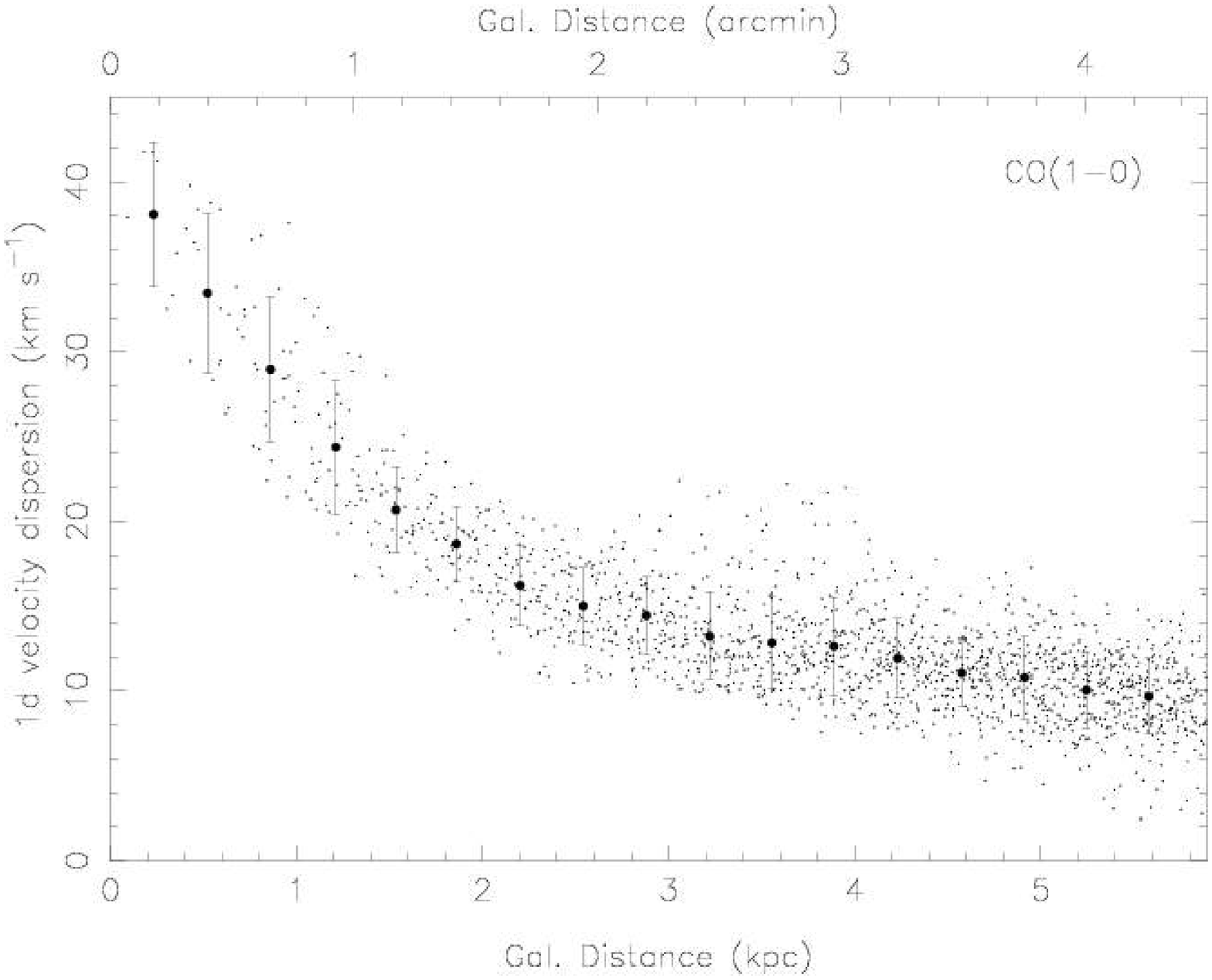}}}
	\resizebox{0.48\hsize}{!}{\rotatebox{-90}
		{\includegraphics{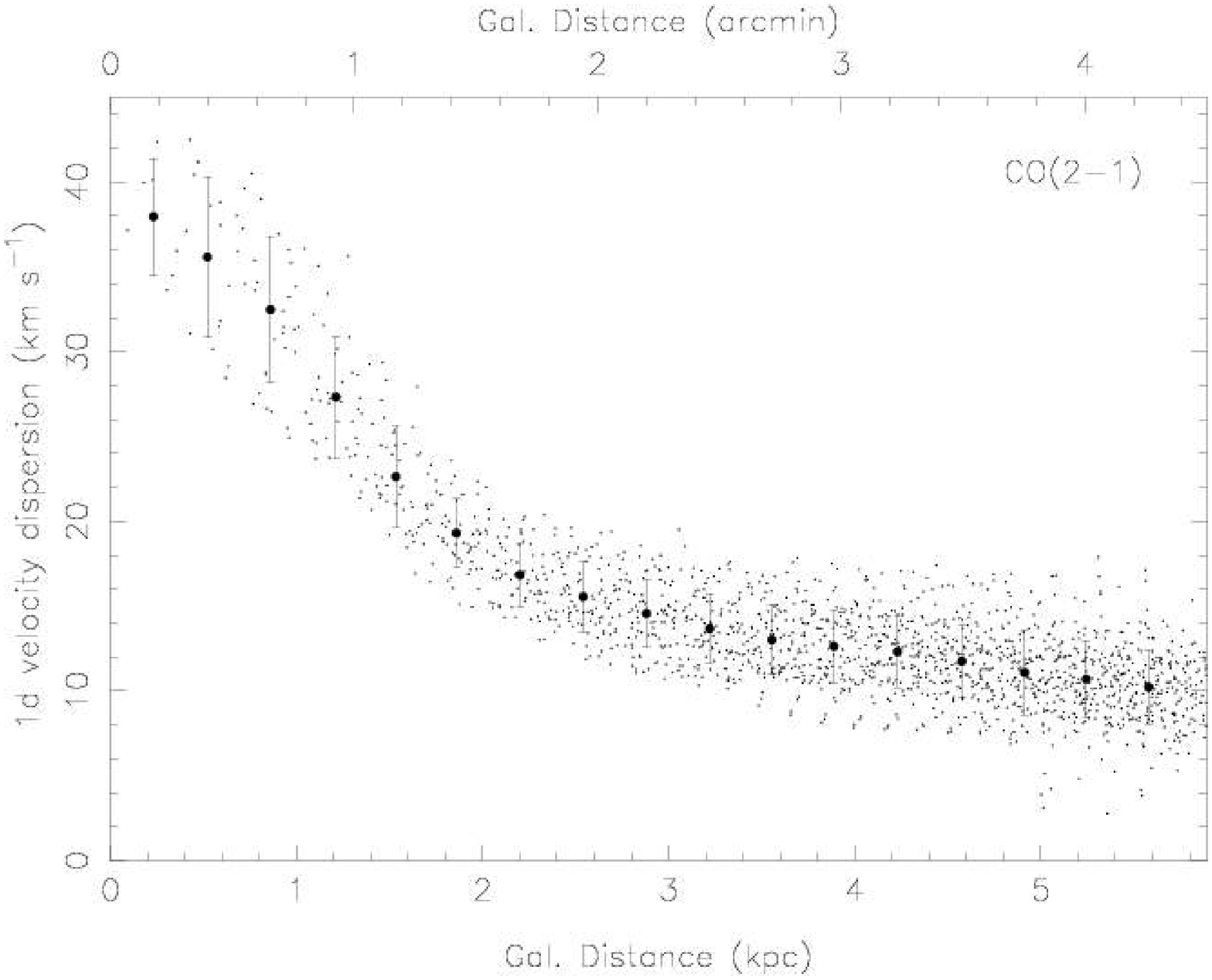}}}
	\caption{Velocity dispersion in the CO(\mbox{$J$=1--0}) and 
	CO(\mbox{$J$=2--1}) 
	data sets, at a common spatial resolution of 49$\arcsec$, as a function of
	the galactocentric distance.
	}
	\label{specwidth}
\end{figure*}

Figure \ref{specwidth} shows the velocity dispersions of the CO($J$=1--0) and 
CO($J$=2--1) lines, at a resolution of 49$\arcsec$, as a function of 
galactocentric radius. In the central region the lines are markedly 
broadened by the steeply rising rotation curve and bar-induced 
non-circular motions. In the disk there is a steady decrease of the 
velocity dispersion with increasing galactocentric radius, reaching a 
value of about 10\,km\,s$^{-1}$ at the edge of the optical disk. The 
dispersion of the CO(\mbox{$J$=2--1}) lines is slightly higher than that of 
the CO(\mbox{$J$=1--0}) lines, by about 0.5--1\,km\,s$^{-1}$ in the disk. This 
could possibly indicate the presence of several cloud components with 
slightly different physical characteristics.

\begin{figure*}
	\resizebox{\hsize}{!}{\rotatebox{0}{
		\includegraphics{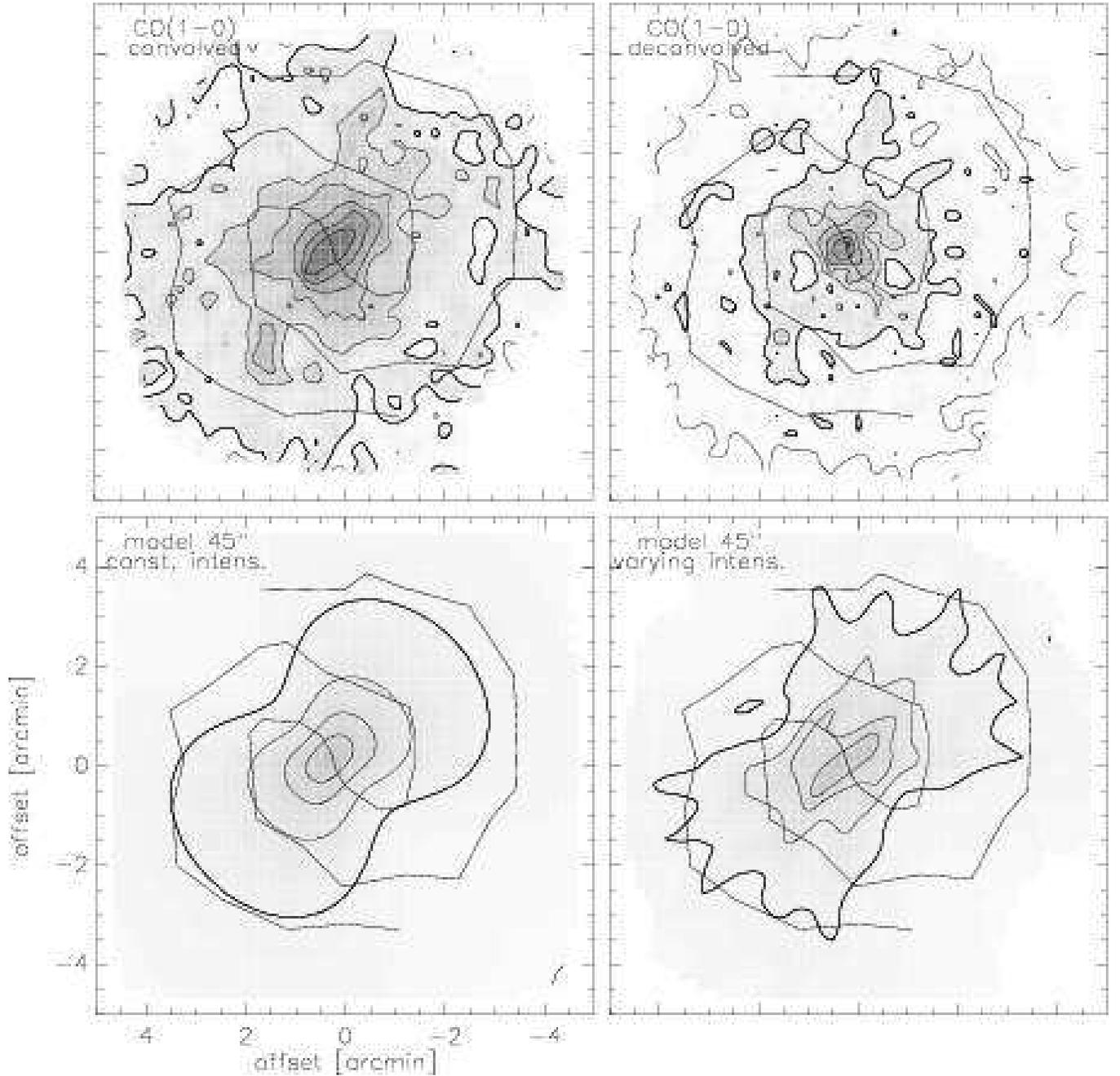}}}
	\caption{Velocity dispersion in the CO(\mbox{$J$=1--0}) data (upper panels),
	two models showing the impact of the effect of the beam 
	coupling to the rotation curve (bottom left) and in addition
	the intensity variation (bottom right). The contours range
	from 5 to 35 km\,s$^{-1}$ and the thick line corresponds to
	15 km\,s$^{-1}$. The increment is 5 km\,s$^{-1}$ in the upper
	panels and 2.5 km\,s$^{-1}$ in the bottom panels (in order
	to enhance the effects in the models).
}
	\label{width}
\end{figure*}

Before taking the decrease of the velocity dispersion in the
disk with increasing galactocentric distance at face value,
we need to evaluate the influence of the beam smearing on
the velocity profiles.
In the convolved CO(\mbox{$J$=1--0}) data set (upper left map in 
Fig.~\ref{width}), the pattern of high velocity dispersion is orthogonal
to the bar, but in the deconvolved data set (upper right map in 
Fig.~\ref{width}) the effect is not present. This suggests
that this feature is only an effect of the beam coupling to 
the intensity distribution in a region where the local velocity and
intensity changes 
within the beam are large.
We made two simple models to show the impact of this effect. 
In the first model we assumed a flat intensity distribution
in order to look at large-scale patterns, and in the second we took 
the intensity distribution from the peak intensity distribution in our 
deconvolved data set in order to show artifacts that may appear on
smaller scales. 
In both models we assumed a fixed velocity dispersion of 8 km 
s$^{-1}$ (see Sect.~\ref{secgma}) and the velocity field generated by the
rotation curve of the CO($J$=1--0) convolved data. 
We created two data cubes, convolved them with a HPBW of 
45$\arcsec$ and calculated the velocity dispersion in the spectra 
using our automated routine (see Paper~I for details).
These maps of the model velocity dispersions can be seen in the bottom 
row of Fig.~\ref{width}. In both cases the velocity dispersion
distribution is
box-shaped and oriented with the ``major axis'' orthogonal to the
bar. At the center the velocity dispersion reaches 20 
km\,s$^{-1}$ in both cases. The model spectra in the central region 
reproduce the shape of the observed spectra nicely. 
However, they underestimate observed velocity dispersion at the 
center (40 km s$^{-1}$), 
which is not surprising given our simple assumptions of the
intensity distribution, rotation curve, and low velocity dispersion
in the nuclear region.
In the model with the varying peak 
intensity, the elongation orthogonal to the bar is more pronounced.
The explanation is that spectra in regions with low levels 
of emission, such as regions close to the bar (along the minor axis), 
are influenced by 
the velocity components in nearby regions with high levels of intensity.
This effect is also present in various regions of the disk. The lack of 
emission between the GMAs (introduced in Sect.~\ref{secgma}) and in the
interarm regions leads here to an artificial broadening of the spectra. 
This results in iso-velocity-dispersion contours with fingers pointing outwards
between the GMAs.

The fact that the artificial broadening of the spectra decreases with
galactocentric radius indicates that beam smearing is responsible for 
at least some of the gradient in the the velocity dispersion 
in Fig.~\ref{specwidth}. 
Our models did not include streaming motions, which also will add to 
the observed velocity dispersion.

\section{Disk instability}\label{secgma}

\begin{figure}
	\resizebox{\hsize}{!}{\rotatebox{-90}{\includegraphics{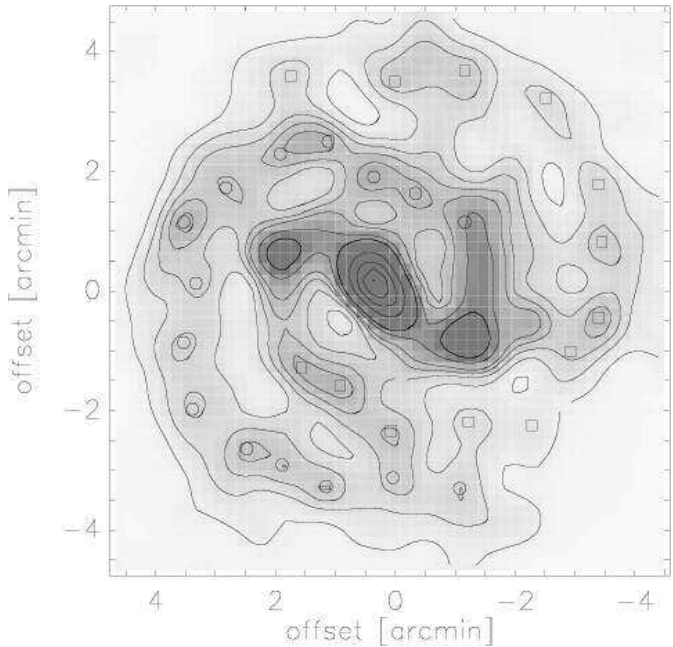}}}
	\caption{Map of the total gas (molecular+atomic) mass surface density
	in M\,83 with
	the position of the GMAs marked with squares and circles for the
	arms originating east and west of the nucleus, respectively.
	The contour values are 10, 20, 30, 40, 50, 75, 100, 200, 300, 400,
	and 500 ${\rm M_\odot~pc^{-2}}$}
 	\label{gmadist}
\end{figure}

Along the spiral arms we find clumpy structures, which
we interpret as Galactic Molecular Associations (GMAs).  
Figure~\ref{gmadist} shows the locations of these structures on a
grey-scale image of the total gas (H$_2$+\ion{H}{i}+He) mass surface density 
($\Sigma_{\rm gas}$). The locations of the GMAs have been estimated from
the velocity-integrated CO(\mbox{$J$=1--0}) emission in the 
{\sc mem}-deconvolved data set. In some ambiguous cases
we have used the peak intensity in the same data set.
The H$_2$ mass surface density is estimated from the
velocity-integrated CO(\mbox{$J$=1--0}) line intensity using
a conversion factor ($X_{\mathrm {CO}}$) of \mbox{2.3 $\times
10^{20}$\,(K\,km\,s$^{-1}$)$^{-1}$\,cm$^{-2}$} (see Paper~I).
The \ion{H}{i} data comes from TA. It was noted 
by TA that due to the lack of short-spacing information 
in the VLA observations only 55\% of the atomic gas mass observed by 
\citet{HB81} with a single dish telescope was detected. In order to calculate 
the total gas mass surface densities, we have to compensate for the missing flux. 
The difference in atomic masses detected in these observations 
corresponds to an average atomic mass surface density of 2.6 
M$_\odot$ pc$^{-2}$. Lacking information on how this mass
is distributed in the galaxy, we take the simplest approach and
add this constant mass surface density to 
the TA data. This is a minor correction, since the typical mass
surface density in the GMAs are of the order 30 M$_\odot$ pc$^{-2}$.
The final data are convolved and regridded to match the sampling
of the CO data. The total gas mass also includes He, through a scaling
of the total \ion{H}{i} and H$_2$ gas mass with 1.36.

The total gas mass of an individual GMA is typically
1--3\,$\times$\,10$^7$ M$_\odot$, and the distance between them are
of the order 1 kpc. Similar structures have been observed in
other galaxies, such as e.g.~M\,51 \citep{KNH95}. 
It is believed that structures like these are formed when the 
gaseous disk becomes unstable to axisymmetric perturbations,
which happens when the mass surface density exceeds a critical density
\citep{T64,KNH95},
\begin{equation}\label{toomre}
	\Sigma_{\rm cr}=\alpha \frac{\sigma_{\rm gas} \kappa}{\pi G}\,,
\end{equation}
where $\alpha$ is a dimensionless factor of order unity, $\sigma_{\rm gas}$ the 
one-dimensional velocity dispersion of the ISM gas, and $\kappa$ the epicyclic
frequency given by 
\begin{equation}
	\kappa^2=\left(R \frac{d\Omega^2}{dR}+4\Omega^2\right)_{R_{\rm g}}\,,
\end{equation}
where $\Omega$ is the angular velocity $v_{\rm c}/R$, and $R_{\rm g}$ the 
center of the epicyclic motion \citep{BT87}.
For a pure stellar or gaseous disk one expects $\alpha$\,=\,1, but in a realistic
gas/stellar disk the value should be $<$\,1 due to the instability introduced 
by the interaction between the two components \citep{JS84,WS94}.
\citet{K89} studied the radial dependence 
of massive star formation and gas mass surface density in a number of galaxies 
and found that $\alpha$ lies in the range 0.5--0.85. An improved study by
\citet{MK01} gave $\alpha$\,=\,0.69 (with some considerable scatter), assuming
a gas velocity dispersion of 6 km s$^{-1}$.
In this paper we use $\alpha$\,=\,0.65.

\begin{figure}
	\resizebox{\hsize}{!}{\rotatebox{-90}{\includegraphics{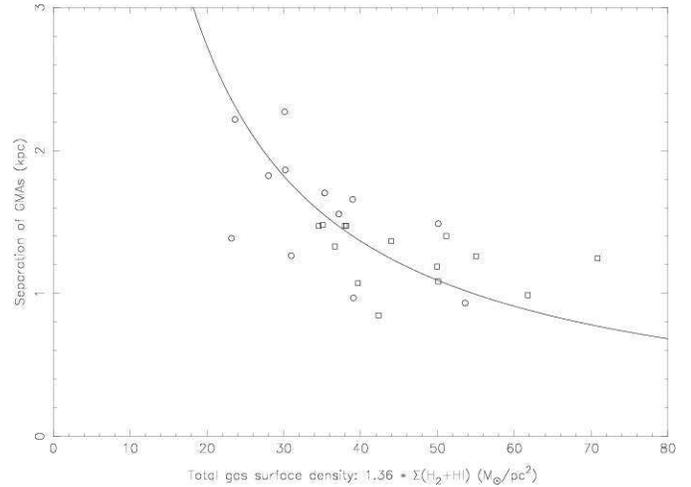}}}
	\caption{The GMA separation as a function of mass surface
	density. The line shows the best fit of (\ref{lambda}) to the data points.
	 }
	\label{gmasep}
\end{figure}

In a gravitationally unstable disk the GMAs are
separated by a distance defined by the wavelength of the fastest growing mode 
in the gaseous disk \citep{E94}. This wavelength can be expressed as
\begin{equation}\label{lambda}
	\lambda=2.2\left(\frac{\sigma_{\rm gas}}{7 {\rm~km~s^{-1}}}\right)^2
	\left(\frac{\Sigma_{\rm gas}}{20 {\rm~M_\odot~pc^{-2}}}\right)^{-1} 
	{\rm~kpc.}
\end{equation}
Figure~\ref{gmasep} shows the separation between the GMAs as a 
function of the average total gas mass surface density in the region. 
Also shown is the best fit of (\ref{lambda}) to the data.
The fit gives an average
velocity dispersion of 7.8 ($\pm 0.9$) km s$^{-1}$, only slightly lower
than the values 9.3 km s$^{-1}$ found for the \ion{H}{i} gas in TA, and 8.0 km s$^{-1}$
for the CO interferometer data in \citet{RLH99}, 
and in agreement with the lower envelope of the values in Fig.~\ref{width}.
This result also gives some credence to our $\Sigma_{\rm gas}$ estimates.

\citet{E94} gives an expression for the characteristic mass of the GMAs 
as a function of velocity dispersion and mass surface density,
\begin{equation}
	M_{\rm GMA}=2.6 \times 10^7\left(\frac{\sigma_{\rm gas}}{7 {\rm~km~s^{-1}}}\right)^4
	\left(\frac{\Sigma_{\rm gas}}{20 {\rm~M_\odot~pc^{-2}}}\right)^{-1} \ {\rm M}_{\odot}.
\end{equation}
This mass agrees well with our GMA mass estimates in M\,83, 1--3\,$\times$\,10$^7$ M$_\odot$.

\begin{figure*}
	\resizebox{\hsize}{!}{\rotatebox{-90}{\includegraphics
	{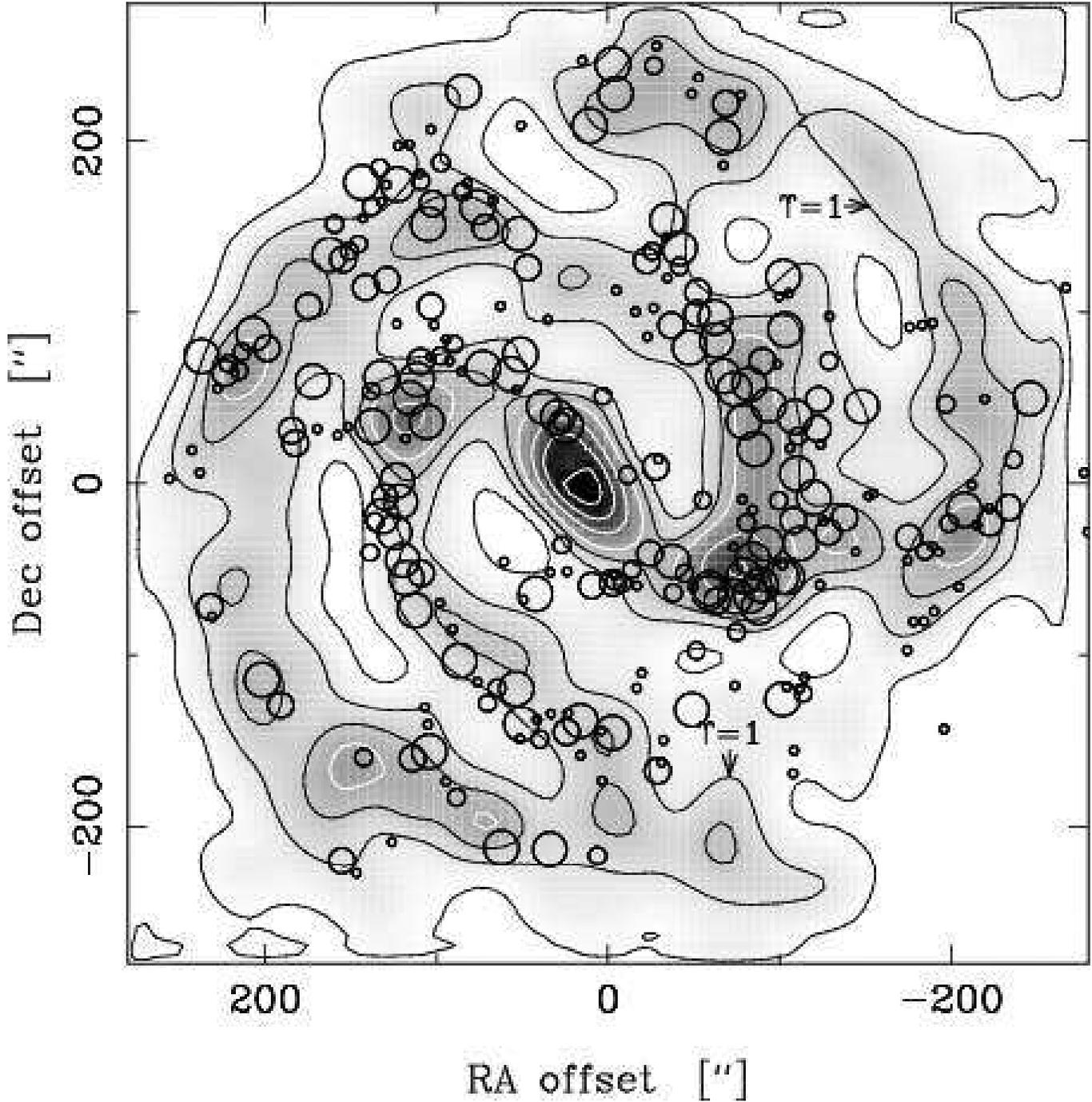}}}
	\caption{A map of the ratio $\Upsilon=\Sigma_{\rm gas}/\Sigma_{cr}$.
	Regions with $\Upsilon$\,$>$\,1 are unstable to axisymmetric perturbations
	and should collapse du to self-gravity of the gas. The contours
	represent $\Upsilon$\,=\,0.5, 1, 1.5, 2, 3, 4, and 5. The first white
	contour is $\Upsilon$\,=\,2. Also shown as circles are the location
	of the \ion{H}{ii} regions in M\,83. The size of a circle reflects the
	luminosity of the \ion{H}{ii} region, ranging from
	non-giant \ion{H}{ii} regions
	(small circles) to giant \ion{H}{ii} regions (large circles). 
	}
	\label{ratio}
\end{figure*}

Using (\ref{toomre}), setting $\sigma_{\rm gas}$\,=\,7.8 km s$^{-1}$, and 
using the rotation curve derived in Sect.~\ref{kinematics}, we have calculated 
the critical mass as a function
of position in M\,83. Figure~\ref{ratio} shows a map of the ratio 
$\Upsilon = \Sigma_{\rm gas}/\Sigma_{\rm cr}$. $\Upsilon$ is below unity between
the arms, hence these regions are stable against gravitational collapse. 
In the arms, $\Upsilon$ is usually about 1--2 and the variation with 
galactocentric distance is small, as expected since star formation drives
$\Upsilon$ towards 1. 
The bar ends and the nucleus show values significantly larger than
1. 
These regions also have massive star formation, but the ratio
is most likely overestimated since the velocity dispersion is higher
than the adopted 7.8 km s$^{-1}$ in these regions (especially 
close to the nucleus). 
Also, the true rotation curve is expected to be steeper than the 
one derived from our CO data, which would give a higher $\kappa$ 
and therefore an increase of the critical density.
The value of $\Upsilon$ correlates nicely with the location
of the \ion{H}{ii}-regions, which are shown as circles in Fig.~\ref{ratio} 
(\ion{H}{ii} data from \citet{RK83}).

\section{Conclusions} We have used CO radio line emission data to investigate the
kinematics of the molecular gas over the entire optical disk
of M83. We compare our data with that of other kinematical
tracers, such as H$\alpha$ and atomic hydrogen.

We find that:

{\it i)} The large scale velocity field is regular. Deviations
from circular rotation in the disk is caused by streaming
motions associated with a spiral density wave. In the center
region the velocity field is strongly coupled to the barred
potential.

{\it ii)} The kinematical center of the CO velocity field coincides
with the IR nucleus of the galaxy. The dynamical mass of the
galaxy within the Holmberg radius is found to be
$6 \times 10^{10}$ M$_{\odot}$. The total gas mass, including
H$_2$, \ion{H}{i} and He, makes up 13\% of this mass.

{\it iii)} The rotation curve is symmetric and adequately fitted
by an exponential mass distribution. The rotation curve
derived from the CO emission agrees with those estimated
from H$\alpha$ and \ion{H}{i} emission.

{\it iv)} Subtracting the global, rotational velocity field we
find that the magnitude and pattern of the residual
velocities of the CO line emission agree with those of \ion{H}{i}.
We also find that the CO residual velocity pattern
bifurcates on the western side. This is expected for
a pattern driven by a spiral density wave with a co-rotation
radius at the galactocentric distance of the bifurcation.
However, no strong bifurcation is found on the eastern side.

{\it v)} Position-velocity maps along the major and minor axes
of M83 show velocity gradients across the spiral arms in
agreement with density wave theory. The velocity shifts
across the spiral arms are 35 km s$^{-1}$ (corrected
for inclination).

{\it vi)} The velocity dispersion in the disk appears to show a 
monotonic decline with galactocentric radii. Part, or all,of 
the decline can be attributed to a convolution of the
velocity field and the finite telescope beam. The velocity
dispersion of the CO(J=2-1) line emission is, however, 
systematically larger than that of the CO(J=1-0) line
emission by 0.5--1 km/s.

{\it vii)} Several Giant Molecular Associations (GMAs) are
identified along the spiral arms of M83. The separation
of these GMAs depends on the velocity dispersion and the
total gas surface density. Our derived CO velocity dispersions
and estimated gas surface densities, together with the observed
GMA separations, give a good fit to the expected  wavelength
of the fastest growing mode of a gravitationally unstable disk.

{\it viii)} The total gas surface mass density, derived from CO
and \ion{H}{i} measurements, is compared to the critical mass surface
density as expected from a Toomre disk. We find that the
observed gas surface mass density exceeds the Toomre value in 
the spiral arm regions, but not in the interarm regions. We
also find that the  number and sizes of \ion{H}{ii} regions correlate
well with the value of the ratio
$\Sigma_{\rm gas}/\Sigma_{\rm cr}$.

\begin{acknowledgements}
We are very grateful toward the SEST staff for their support
during observations and Swedish Natural Science Research Council
for travel expenses support.
We also wish to thank Remo Tilanus and Ron Allen
for letting us use their \ion{H}{i} data.
% and S\"oren Larsen for the use of the
%optical images and the referee for valuable input. 
\end{acknowledgements}

\bibliography{H0230ref}
\bibliographystyle{aa}

\end{document}